\documentclass[prb,twocolumn]{revtex4}
\usepackage{graphicx}

\begin{document}

\title{
%Theory of superconducting Fermi liquids without inversion symmetry
Fermi liquid theory 
for heavy fermion superconductors without inversion symmetry
: Magnetism and transport coefficients
}

\author{Satoshi Fujimoto} 
\affiliation{
Department of Physics,
Kyoto University, Kyoto 606-8502, Japan
}

\date{\today}

\begin{abstract}
We present the microscopic Fermi liquid theory for 
%general theoretical descriptions for 
magnetic properties and transport phenomena
in interacting electron systems without inversion symmetry
both in the normal state and in the superconducting state.
%on the basis of the Fermi liquid theory.
Our argument is mainly focused on the application to
noncentrosymmetric heavy fermion superconductors.
% such as CePt$_3$Si,
%CeRhSi$_3$, and CeIrSi$_3$.
The transport coefficients for
the anomalous Hall effect, the thermal anomalous Hall effect, 
the spin Hall effect, and magnetoelectric effects, 
of which the existence is a remarkable feature
of parity violation,
are obtained by taking into account electron correlation effects in a 
formally exact way. 
%It is found that the spin Hall conductivity is not renormalized
%by electron correlation effects, and is determined only by
%the band structure.
%It is pointed out that 
%the thermal anomalous Hall conductivity 
%in the superconducting state
%behaves like in the normal state even in the zero temperature limit,
%provided that 
%the spin-orbit splitting is much larger than the superconducting gap. 
Moreover, we demonstrate that 
the temperature dependence of
the spin susceptibility which consists of
the Pauli term and van-Vleck-like term seriously depends
on the details of the electronic structure.
We give a possible explanation for
the recent experimental result obtained by the NMR measurement 
of CePt$_3$Si [Yogi et al.: J. Phys. Soc. Jpn.{\bf 75} (2006) 013709], 
which indicates no change of the Knight shift below 
the superconducting transition temperature 
for any directions of a magnetic field.
%we propose a possible solution
%to 
%A characteristic magnetic property of 
%In the superconducting state, the van Vleck term of
%the uniform spin susceptibility caused
\end{abstract}

\maketitle
%%%%%
%%%%%%%%%%%%%%%%%%%%%%%%%%%%%%%%%%%%%%%%%%%%%%%%%
\section{Introduction}

\begin{figure}
\begin{center}
\includegraphics[width=8cm]{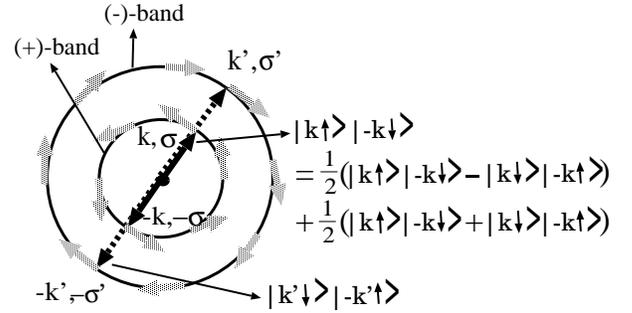}
\end{center}
\caption{An example of 
two-dimensional Fermi surfaces splitted by the Rashba
spin-orbit interaction depicted on the $k_x$-$k_y$ plane.
The inversion symmetry is broken along the $z$-direction. 
The gray arrows represent electron spins on the Fermi surfaces for
the $(+)$-band and the $(-)$-band.
The Cooper pairs between electrons with momentum $k$, spin $\sigma$ 
and momentum $-k$, spin $-\sigma$ are formed on each
Fermi surfaces. }
\label{fig1}
\end{figure}

For a large class of superconductors, 
there exists inversion symmetry which 
allows the classification of
the Cooper pair according to parity; i.e.
a spin singlet or triplet pairing state realizes. 
This idea is not applicable to 
the recently discovered noncentrosymmetric
superconductors,
CePt$_3$Si, UIr, CeRhSi$_3$, CeIrSi$_3$, and 
Li$_2$(Pd$_{1-x}$Pt$_x$)B.\cite{bau,uir,kim,onuki,tog1,tog2}
In systems without inversion center,
parity violation yields the admixture of
the spin singlet and triplet pairing states, as pointed out by Edelstein
nearly two decades ago.\cite{ede1,gor}
An asymmetric potential gradient $\nabla V$ caused by
the absence of inversion symmetry gives rise to
the spin-orbit interaction $\alpha (\mbox{\boldmath $k$}\times \nabla V)
\cdot\mbox{\boldmath $\sigma$}$ which splits the Fermi surface
into two pieces and aligns electron spins on each Fermi surfaces parallel to
$\mbox{\boldmath $k$}_F\times\nabla V$, with $\mbox{\boldmath $k$}_F$ 
the Fermi momentum.
A simple example of this situation
for the Rashba type spin-orbit interaction with
$\nabla V$ parallel to the $z$-axis 
is depicted in FIG.\ref{fig1}.
Note that in FIG.1 the spin quantization axis is chosen
along $\mbox{\boldmath $k$}_F\times\nabla V$.
On one of the spin-orbit splitted Fermi surfaces,
say the $(+)$-band in FIG.\ref{fig1},
the Cooper pair between electrons with momentum $k$, spin $\uparrow$
and momentum $-k$, spin $\downarrow$ is formed. 
This state, denoted as $|k\uparrow\rangle |-k\downarrow\rangle$,
is {\it not} a spin singlet state, because
the counterpart of this state $|k\downarrow\rangle |-k\uparrow\rangle$
is formed on another Fermi surface and
thus the superposition between these two states
is not possible.
Actually, the pairing state $|k\uparrow\rangle |-k\downarrow\rangle$ 
is the admixture of spin singlet and triplet states as easily verified by,
\begin{eqnarray}
|k\uparrow\rangle |-k\downarrow\rangle&=&
\frac{1}{2}(|k\uparrow\rangle |-k\downarrow\rangle -
|k\downarrow\rangle |-k\uparrow\rangle) \nonumber \\
&+&
\frac{1}{2}(|k\uparrow\rangle |-k\downarrow\rangle +
|k\downarrow\rangle |-k\uparrow\rangle). \nonumber 
\end{eqnarray}
The first and second terms
of the right-hand side express, respectively, 
a spin singlet state and 
a spin triplet state with the in-plane spin projection $S_{\rm in plane}$ 
equal to $0$.
Since we take the spin quantization axis parallel to the $xy$-plane,
this triplet state corresponds to the $S^z=\pm 1$ state
for the spin quantization axis along the $z$-direction.
(This means that the $\vec{d}$-vector of the triplet component
is parallel to the plane.)
The above explanation 
is also applicable to general cases with more complicated
form of $\nabla V$.
This unique superconducting state exhibits
various interesting electromagnetic properties
as extensively argued by many 
authors.\cite{ede1,gor,fri,ser,ede2,yip,yip2,sam2,kau,fuji2,haya,tana,ichi}

In addition to the superconducting state,
the absence of inversion symmetry also affects properties of
the normal state in a drastic way.
One remarkable feature appears in paramagnetic effects;
i.e. there is a van-Vleck-like spin susceptibility
due to the transition between the spin-orbit splitted Fermi 
surfaces.\cite{ede1,gor,fri} 
This effect should be distinguished from the usual van Vleck term
of the orbital susceptibility by the fact
that in contrast to the usual van Vleck term,
% which is independent of temperature, 
the van-Vleck-like susceptibility
in inversion-symmetry-broken systems varies
as a function of temperature like the Pauli susceptibility.
The existence of the van-Vleck-like term is crucial 
for experimental investigations on superconducting states
because it exhibits no change even below 
the superconducting transition temperature $T_c$
for both the spin-singlet and triplet states
when the spin-orbit splitting is sufficiently
larger than the superconducting gap.\cite{gor,fri}
For the determination of the symmetry of the Cooper pair,
we need precise knowledge about the behaviors of the van-Vleck-like 
susceptibility both below and above $T_c$.

Furthermore, 
the parity-breaking spin-orbit interaction gives rise to
nontrivial coupling between charge and spin degrees of freedom,
and thus causes unique transport properties.
For example, there exist magnetoelectric effects such as
the electric-field-induced magnetization,\cite{lev,ede0}
and the spin Hall effect which is characterized by
the transverse spin current caused by an electric field.\cite{she,she2}
Also, the anomalous Hall effect
considered by Karplus and Luttinger five decades ago is 
possible to occur.\cite{kl}
For heavy fermion systems without inversion symmetry, 
the experimental studies on these effects are important future issues.

In the heavy fermion superconductors
CePt$_3$Si, CeRhSi$_3$, UIr, and CeIrSi$_3$, 
strong electron 
correlation which yields an extremely enhanced effective electron mass
plays a crucial role.
For the clarification of the nature of
these systems,
it is important to understand 
effects of the electron correlation on the aforementioned exotic phenomena. 
The main part of this paper is devoted to this purpose.
On the basis of a general framework of
the Fermi liquid theory, we derive the formulae for
the spin susceptibility and the transport coefficients related to
the anomalous Hall effect, the thermal anomalous Hall effect, the spin Hall
effect and the magnetoelectric effect taking into account 
strong electron correlation effects in a formally exact way. 
Using these formulae, we argue how strong electron correlation inherent
in heavy fermion systems affect experimentally
observable quantities. 
The insights obtained from this analysis are useful 
for the exploration of the superconducting state
realized in CePt$_3$Si, CeRhSi$_3$, and CeIrSi$_3$, as explained below.

The main results of this paper are summarized as follows:

(i) The thermal anomalous Hall effect, which is the anomalous Hall
effect for  the heat current is predominated by
the contributions from electrons away from the Fermi surface.
It implies that when the spin-orbit splitting is much larger than the
superconducting gap, the behavior of the thermal anomalous Hall conductivity 
$\kappa^{\rm AHE}_{xy}$ 
in the superconducting state is similar to
that in the normal state.
Even in the zero temperature limit $T\rightarrow 0$, 
$\kappa^{\rm AHE}/TH_z$ with $H_z$ an applied magnetic field
is finite, and much enhanced by electron correlation effects 
in heavy fermion systems.

(ii) In the Rashba case where the inversion symmetry is broken along the 
$z$-direction, as in the case of CePt$_3$Si, CeRhSi$_3$, and
 CeIrSi$_3$, 
the spin Hall conductivity, which is given by
the correlation function between a charge current and a transverse
spin current, is not at all affected by electron correlation effects,
but determined only by the band structure.

(iii) We clarify electron correlation effects on the magnetoelectric 
effect both in the normal state and in the superconducting state.
As was considered by Levitov {\it et al.},\cite{lev} in the normal state,
the bulk magnetization is induced by an applied electric field,
$M=\Upsilon E$,
and, conversely, the charge current flow is induced by an AC magnetic field,
$J=-\Upsilon \dot{B}$.
We find that in the temperature region where the resistivity
exhibits $T^2$-dependence, the magnetoelectric effect coefficient
$\Upsilon$ 
behaves like $\sim 1/(\gamma T^2)$, 
with $\gamma$ the specific heat coefficient,
while in the temperature region where the resistivity is dominated 
by impurity scattering, $\Upsilon$ is enhanced by a factor $\sim \gamma$.
In the superconducting state, 
the paramagnetic supercurrent is generated by the Zeeman field,
as extensively argued by several authors.\cite{ede1,ede2,yip,fuji2}
We reveal electron correlation effects on this supercurrent
from a general theoretical framework.

(iv) The temperature dependence of 
the van-Vleck-like spin susceptibility $\chi^{\rm VV}(T)$
%, which exists because
%of the inversion-symmetry-breaking spin-orbit interaction, 
crucially depends on the details of the electronic structure, 
and occasionally may become 
stronger than that of the Pauli susceptibility $\chi^{\rm Pauli}(T)$,
in contrast to the usual van Vleck orbital susceptibility, which is 
independent of temperatures.
Also, in the superconducting state, 
the ratio of $\chi^{\rm VV}(T)$ to the total spin susceptibility 
$\chi(T)=\chi^{\rm Pauli}(T)+\chi^{\rm VV}(T)$ 
in the zero temperature limit is seriously affected by
both the electronic structure and electron correlation effects.
This result yields an important implication for
the recent NMR measurements for CePt$_3$Si carried out by 
Yogi {\it et al.},\cite{yogi,yogi2} 
which indicate no change of the Knight shift below $T_c$ for any directions
of an applied magnetic field, contrary to
the previous theoretical prediction that for a magnetic field
along  the $z$-axis, $\chi(0)=\chi(T_c)/2$ for 
a sufficiently strong spin-orbit coupling.
Our precise analysis indicates that for a specific 
electronic structure, $\chi(T)$ is dominated by $\chi^{\rm VV}$, and
as a result, $\chi(0)\approx \chi(T_c)$, which is consistent with
the aforementioned NMR experiment.

The organization of this paper is as follows.
In the next section, we develop the analysis for the normal state,
exploring the static magnetic properties and the transport coefficients. 
In the section III, we extend our approach to the superconducting
with particular emphasis on the magnetism.
We propose a possible explanation for the recent NMR experimental data.
We also argue electron correlation effects on
the paramagnetic supercurrent caused by
the inversion-symmetry-breaking spin-orbit interaction.
A summary is given in the section IV.

\section{Normal state}

\subsection{Basic equations}

In this section,
we consider a single band interacting electron model with a lattice structure
breaking inversion symmetry.
An extension to more complicated systems such as the periodic Anderson model
is straightforward, and will be discussed elsewhere. 
Our model Hamiltonian is given by, 
\begin{eqnarray}
\mathcal{H}&=&\mathcal{H}_0+\mathcal{H}_{\rm SO}, \label{ham} \\
\mathcal{H}_0&=&\sum_{k,\sigma} \varepsilon_k c^{\dagger}_{k}c_{k}
+U\sum_i n_{\uparrow i}n_{\downarrow i}, \label{ham0} \\
\mathcal{H}_{\rm SO}&=&\alpha\sum_{k}
c^{\dagger}_{k}\mbox{\boldmath $\mathcal{L}$}_0(k)
\cdot\mbox{\boldmath $\sigma$}c_{k},
%\mathcal{H}_0&=&\sum_{k,\sigma} 
%\varepsilon_k c^{\dagger}_{\sigma k}c_{\sigma k}
%+U\sum_i n_{\uparrow i}n_{\downarrow i} \\
%\mathcal{H}_{\rm SO}&=&\alpha\sum_{k,\sigma\sigma'}
%\mbox{\boldmath $\mathcal{L}$}_0(k)
%\cdot\mbox{\boldmath $\sigma$}_{\sigma\sigma'}
%c^{\dagger}_{\sigma k}c_{\sigma' k},
\label{so}
\end{eqnarray}
where $c^{\dagger}_k=(c^{\dagger}_{\uparrow k},
c^{\dagger}_{\downarrow k})$ 
is the two-component
spinor field for an electron with spin $\uparrow$, $\downarrow$,
and momentum $k$.
$n_{\sigma i}=c^{\dagger}_{\sigma i}c_{\sigma i}$ is 
the number density operator at the site $i$. 
$\mbox{\boldmath $\sigma$}=(\sigma_x,\sigma_y,\sigma_z)$ 
with $\sigma_{\nu}$, $\nu=x,y,z$, the Pauli matrix.
For simplicity, we have assumed the on-site Coulomb repulsion
with the coupling constant $U$ in Eq.(\ref{ham0}).
The vector $\mbox{\boldmath $\mathcal{L}$}_0(k)
=({\cal L}_{0x},{\cal L}_{0y},{\cal L}_{0z})$
in the spin-orbit interaction term $\mathcal{H}_{\rm SO}$
obeys $\mbox{\boldmath $\mathcal{L}$}_0(-k)
=-\mbox{\boldmath $\mathcal{L}$}_0(k)$, 
of which the explicit expression is determined 
by the detail of the crystal structure which breaks the inversion symmetry.
For the tetragonal symmetry and small $k$, 
$\mbox{\boldmath $\mathcal{L}$}_{0}(k)=(k_y,-k_x,0)$
which leads the Rashba-type spin-orbit interaction.\cite{ras}
For the cubic symmetry with Zinc Blende structures and small $k$, 
$\mbox{\boldmath $\mathcal{L}$}_{0}(k)=(k_x(k_y^2-k_z^2),
k_y(k_z^2-k_x^2),k_z(k_x^2-k_y^2))$, which is related to the Dresselhaus
spin-orbit interaction.\cite{dres,dyp,sil}

In the following, we are concerned with magnetic properties of the system.
Unique features of magnetism in inversion-symmetry-broken systems appear
as a result of paramagnetic effects rather than diamagnetic effects.
Thus we add to the Hamiltonian (\ref{ham}) the Zeeman coupling with an
external magnetic field $\mbox{\boldmath $H$}=(H_x,H_y,H_z)$,
\begin{eqnarray}
H_{\rm Zeeman}=\sum_kc^{\dagger}_k
\mu_{\rm B}\mbox{\boldmath $\sigma$}\cdot\mbox{\boldmath $H$}
c_k, \label{zee}
\end{eqnarray} 
and neglect the orbital diamagnetic effects.
For simplicity, we assume that the $g$-value is equal to 2.
Then, the inverse of the single-particle
Green's function under the applied magnetic field 
is defined as,
\begin{eqnarray}
\hat{G}^{-1}(p)=i\varepsilon_n-\hat{H}(p)+
\mu_{\rm B}\mbox{\boldmath $\sigma$}\cdot\mbox{\boldmath $H$}
, \label{h0} 
\end{eqnarray}
where $p=(i\varepsilon_n,\mbox{\boldmath $k$})$, and,
\begin{eqnarray}
\hat{H}(p)=\hat{H}_0(p)+\hat{\Sigma}(p), \label{h1}
\end{eqnarray}
\begin{eqnarray}
\hat{H}_0=\varepsilon_k-\mu+
\alpha\mbox{\boldmath $\mathcal{L}$}_0(k)\cdot\mbox{\boldmath $\sigma$}
\label{h12}
\end{eqnarray}
with $\mu$ the chemical potential. 
The self-energy matrix $\hat{\Sigma}$ consists of 
both diagonal and off-diagonal components,
\begin{eqnarray}
\hat{\Sigma}&=&
\left(
\begin{array}{cc}
 \Sigma_{\uparrow\uparrow}(p)   & \Sigma_{\uparrow\downarrow}(p)  \\
 \Sigma_{\downarrow\uparrow}(p) & \Sigma_{\downarrow\downarrow}(p)
\end{array}
\right)   \nonumber \\
&=&\Sigma_0+\mbox{\boldmath $\Sigma$}
\cdot\mbox{\boldmath $\sigma$}. \label{self}
\end{eqnarray}
Here $\mbox{\boldmath $\Sigma$}=(\Sigma_x,\Sigma_y,\Sigma_z)$ 
with $\Sigma_0=(\Sigma_{\uparrow\uparrow}+\Sigma_{\downarrow\downarrow})/2$,
$\Sigma_x=(\Sigma_{\downarrow\uparrow}+\Sigma_{\uparrow\downarrow})/2$, 
$\Sigma_y=(\Sigma_{\downarrow\uparrow}-\Sigma_{\uparrow\downarrow})/2i$, and 
$\Sigma_z=(\Sigma_{\uparrow\uparrow}-\Sigma_{\downarrow\downarrow})/2$.
%Since $\hat{\Sigma}(p)$ is Hermitian,
%$\Sigma_{\downarrow\uparrow}(\vec{p},z)=
%\Sigma_{\uparrow\downarrow}^{*}(\vec{p},z)$. 
%In the absence of magnetic fields $\vec{H}=0$, 
%$\Sigma_0=
%\Sigma_{\uparrow\uparrow}=\Sigma_{\downarrow\downarrow}\equiv \Sigma$

It is convenient to introduce a vectorial function,
\begin{eqnarray}
\mbox{\boldmath $\mathcal{L}$}(p)=({\cal L}_x(p),{\cal L}_y(p),{\cal L}_z(p))
=\mbox{\boldmath $\mathcal{L}$}_{0}(k)
-\frac{\mu_{\rm B}}{\alpha}\mbox{\boldmath $H$}+\frac{1}{\alpha}
\mbox{\boldmath $\Sigma$}(p).
\end{eqnarray}
Generally, $\mbox{\boldmath $\mathcal{L}$}(p)$ is non-Hermitian, because 
$\Sigma_{x}$ and $\Sigma_y$ may be complex quantities.
However, 
%Then, 
$\hat{G}^{-1}(p)=i\varepsilon-\varepsilon_k+\mu-\Sigma_0(p)
-\alpha\mbox{\boldmath $\mathcal{L}$}(p)\cdot\mbox{\boldmath $\sigma$}$ 
is diagonalized by the transformation 
%$\hat{U}(p)\hat{G}^{-1}(p)\hat{U}^{\dagger}(p)$
$\hat{U}(p)\hat{G}^{-1}(p)\hat{U}_{+}(p)$
with,
\begin{eqnarray}
\hat{U}(p)=
\left(
\begin{array}{cc}
\xi_{+}(p)  &   \xi_{-}(p)\eta_{-}(p) \\
-\xi_{-}(p)\eta_{+}(p) & \xi_{+}(p) 
\end{array}
\right), \label{unita}
\end{eqnarray}
\begin{eqnarray}
\hat{U}_{+}(p)=
\left(
\begin{array}{cc}
\xi_{+}(p)  &   -\xi_{-}(p)\eta_{-}(p) \\
\xi_{-}(p)\eta_{+}(p) & \xi_{+}(p) 
\end{array}
\right), \label{unita2}
\end{eqnarray}
\begin{eqnarray}
\xi_{\pm}(p)=\frac{1}{\sqrt{2}}\left[1\pm\frac{{\cal L}_z(p)}
{\Vert\mbox{\boldmath $\mathcal{L}$}(p)\Vert}
\right]^{\frac{1}{2}},
\end{eqnarray}
\begin{eqnarray}
\eta_{\pm}(p)=\frac{{\cal L}_x(p)\pm i{\cal L}_y(p)}
{\sqrt{{{\cal L}_x(p)}^2+{{\cal L}_y(p)}^2}}.
%
%\hat{U}(p)=
%\left(
%\begin{array}{cc}
%\xi_{+}(p)  &   \xi_{-}(p)e^{-i\theta_p} \\
%-\xi_{-}(p)e^{i\theta_p} & \xi_{+}(p) 
%\end{array}
%\right), \label{unita}
%\end{eqnarray}
%\begin{eqnarray}
%\xi_{\pm}(p)=\frac{1}{\sqrt{2}}\left[1\pm\frac{{\cal L}_z(p)}
%{|\mbox{\boldmath $\mathcal{L}$}(p)|}
%\right]^{\frac{1}{2}},
%\end{eqnarray}
%\begin{eqnarray}
%e^{i\theta_p}=\frac{{\cal L}_x(p)+i{\cal L}_y(p)}
%{\sqrt{{{\cal L}_x(p)}^2+{{\cal L}_y(p)}^2}}
\end{eqnarray}
Here $\Vert\mbox{\boldmath $\mathcal{L}$}(p)\Vert=
\sqrt{\mathcal{L}_x^2+\mathcal{L}_y^2+\mathcal{L}_z^2}$.
In the absence of the magnetic field,
we can verify that the time reversal symmetry leads
the relation 
\begin{eqnarray}
\Sigma_{\uparrow\downarrow}^{*}(z,-\mbox{\boldmath $k$})
=-\Sigma_{\downarrow\uparrow}(z,\mbox{\boldmath $k$}).
\label{rel1}
\end{eqnarray}

As seen from eqs.(\ref{h1}),(\ref{h12}), and (\ref{self}),
the main effect of the off-diagonal self-energy terms, 
$\Sigma_{\uparrow\downarrow}$, $\Sigma_{\downarrow\uparrow}$ 
is to renormalize
the spin-orbit interaction term.
Since the on-site Coulomb interaction does not change 
the symmetry of the system, the off-diagonal self-energy
should obey the same symmetric properties as
$\mbox{\boldmath $\mathcal{L}$}_0(k)$
in the absence of magnetic fields; i.e.
\begin{eqnarray}
\mbox{\boldmath $\Sigma$}(i\varepsilon_n,-\mbox{\boldmath $k$})=
-\mbox{\boldmath $\Sigma$}(i\varepsilon_n,\mbox{\boldmath $k$}).
\label{rel2}
\end{eqnarray}
In particular, for the Rashba interaction,
\begin{eqnarray}
\Sigma_{x}(k_x,-k_y)=
-\Sigma_{x}(k_x,k_y),
\end{eqnarray}
\begin{eqnarray}
\Sigma_{y}(-k_x,k_y)=
-\Sigma_{y}(k_x,k_y),
\end{eqnarray}
and $\Sigma_{x}$ 
($\Sigma_{y}$) is 
an even function of $k_x$ ($k_y$). 

The relations (\ref{rel1}) and (\ref{rel2}) imply that
when the time reversal symmetry is preserved, 
$\Sigma_{x}$ and $\Sigma_{y}$ are real quantities, and thus
$\hat{U}(p)$ is unitary,
i.e. $\hat{U}_{+}(p)=\hat{U}^{\dagger}(p)$.
In more general cases with $\mbox{\boldmath $H$}\neq 0$,
as long as the spin-orbit splitting is much smaller than the Fermi energy, 
as in the case of any heavy fermion systems without inversion symmetry,
the off-diagonal terms of $\hat{\Sigma}$ may be negligible
compared to the diagonal terms, and
$\hat{U}_{+}$ is safely approximated by $\hat{U}^{\dagger}$.

The single-particle excitation energy $\varepsilon_{k\tau}^{*}$ 
for the quasi-particle with the helicity $\tau=\pm 1$ 
is given by the solution of the equation 
${\rm Det}[\hat{G}^{-1}(\varepsilon^{*}_{k\tau},\mbox{\boldmath $k$})]=0$,
which is, in the diagonalized representation,
\begin{equation}
\varepsilon_{k\tau}^{*}+\mu-\varepsilon_{k}
-\Sigma_{0}(\varepsilon_{\tau}^{*},\mbox{\boldmath $k$})
-\tau\alpha \Vert\mbox{\boldmath $\mathcal{L}$}
(\varepsilon_{k\tau}^{*},\mbox{\boldmath $k$})
\Vert=0.
\end{equation}
Using the transformation $\hat{U}(p)$,
we can easily obtain the inverse of Eq.(\ref{h0}),
\begin{eqnarray}
\hat{G}(p)=\sum_{\tau=\pm 1}\frac{1+\tau
\hat{\mbox{\boldmath $\mathcal{L}$}}(p)\cdot
\mbox{\boldmath $\sigma$}}{2}G_{\tau}(p), \label{gree}
\end{eqnarray}
where $\hat{\mbox{\boldmath $\mathcal{L}$}}(p)=
\mbox{\boldmath $\mathcal{L}$}(p)/
\Vert\mbox{\boldmath $\mathcal{L}$}(p)\Vert$,
and 
\begin{eqnarray}
G_{\tau}(p)=\frac{1}{i\varepsilon_n-\varepsilon_k+\mu-\Sigma_{0}(p)
-\tau\alpha \Vert\mbox{\boldmath $\mathcal{L}$}(p)\Vert}. \label{grepm}
\end{eqnarray}
In the vicinity of the Fermi surface, the quasiparticle approximation
is applicable, 
and the Green function for $\mbox{\boldmath $H$}=0$ is reduced to,
\begin{eqnarray}
\hat{G}(p)=\sum_{\tau=\pm 1}\frac{1+\tau
\hat{\mbox{\boldmath $\mathcal{L}$}}
(\varepsilon_{k\tau}^{*},\mbox{\boldmath $k$})
\cdot\mbox{\boldmath $\sigma$}}{2}G_{\tau}(p),
\end{eqnarray}
with 
\begin{eqnarray}
G_{\tau}(p)=\frac{z_{k\tau}}{i\varepsilon_n
+i\gamma_{k\tau} {\rm sgn}\varepsilon_n
-\varepsilon_{k\tau}^{*}}
\end{eqnarray}
Here $\gamma_{k\tau}$ is the quasi-particle damping given by
\begin{eqnarray}
\gamma_{k\tau}=z_{k\tau}({\rm Im}\Sigma_0^{R}
(\varepsilon,\mbox{\boldmath $k$})
+\alpha{\rm Re}\hat{\mbox{\boldmath $\mathcal{L}$}}^R
%\frac{{\rm Re}\mbox{\boldmath $\mathcal{L}$}^R
%(\varepsilon,\mbox{\boldmath $k$})}{|{\rm Re}\mbox{\boldmath $\mathcal{L}$}^R
%(\varepsilon,\mbox{\boldmath $k$})|}
\cdot{\rm Im}\mbox{\boldmath $\Sigma$}^R
(\varepsilon,\mbox{\boldmath $k$})
)|_{\varepsilon\rightarrow i\varepsilon_n},
\label{damp}
\end{eqnarray}
where ${\rm Re}\hat{\mbox{\boldmath $\mathcal{L}$}}^R=
{\rm Re}\mbox{\boldmath $\mathcal{L}$}^R
(\varepsilon,\mbox{\boldmath $k$})/|{\rm Re}\mbox{\boldmath $\mathcal{L}$}^R
(\varepsilon,\mbox{\boldmath $k$})|$
and quantities with the superscript $R$ indicates 
retarded functions obtained by
the analytic continuation $i\varepsilon_n\rightarrow \varepsilon+i\delta$ 
($\delta>0$).
%$\gamma_{k\tau}=z_{k\tau}{\rm Im}\Sigma_0(k,\varepsilon_{k\tau}^{*})$, 
The mass renormalization factor $z_{k\tau}$ is
\begin{eqnarray}
z_{k\tau}=\biggl[1-\frac{\partial \Sigma_0(p)}
{\partial (i\varepsilon_n)} 
-\tau \alpha \frac{\partial \Vert\mbox{\boldmath $\mathcal{L}$}(p)\Vert}
{\partial (i\varepsilon_n)}
\biggr]^{-1}\biggr|_{i\varepsilon_n\rightarrow\varepsilon^{*}_{k\tau}}.
\label{zkt}
\end{eqnarray}
In particular, 
in the case of the Rashba interaction with
 a potential gradient 
along $\mbox{\boldmath $n$}=(0,0,1)$,
\begin{eqnarray}
z_{k\tau}&=&\biggl[1-\frac{\partial \Sigma_0(p)}
{\partial (i\varepsilon_n)} 
-\tau \biggl(\hat{t}_y\frac{\partial \Sigma_x(p)}{\partial (i\varepsilon_n)}
\nonumber \\
&-&\hat{t}_x\frac{\partial \Sigma_y(p)}{\partial (i\varepsilon_n)}\biggr) 
\biggr]^{-1}\biggr|_{i\varepsilon_n\rightarrow\varepsilon^{*}_{k\tau}},
\label{rz}
\end{eqnarray}
%%%%%%%%%%%%%amended
with a two-dimensional vector $\hat{\mbox{\boldmath $t$}}(p)=
\mbox{\boldmath $t$}(p)/\Vert\mbox{\boldmath $t$}(p)\Vert$,
 $\mbox{\boldmath $t$}(p)=\mbox{\boldmath $n$}\times
\mbox{\boldmath {$\mathcal{L}$}}(p)$, and 
$\Vert\mbox{\boldmath $t$}(p)\Vert=\sqrt{t_x^2+t_y^2}$.
In the derivation of Eqs.(\ref{damp}), (\ref{zkt}), and (\ref{rz}),
it is assumed that the quasiparticle damping is sufficiently small
to stabilize the Fermi liquid state; i.e. 
${\rm Im}\Sigma_{\mu}^R(\varepsilon,\mbox{\boldmath $k$})\ll \varepsilon$ 
with $\mu=0,x,y,z$.
We would like to stress that the notations $\Sigma_{x,y}$
is more convenient than the notations 
$\Sigma_{\uparrow\downarrow}$ and $\Sigma_{\downarrow\uparrow}$,
because ${\rm Im}\Sigma^R_{\uparrow\downarrow}$ and  
${\rm Im}\Sigma^R_{\downarrow\uparrow}$
are not directly related to the quasiparticle damping, 
and can not be assumed to be small.  
%%%%%%%%%%%%%%%%%%%%%%%

\subsection{Ward identities}

In the following sections, we will argue 
transport phenomena and magnetic properties.
In the derivation of transport coefficients, the Ward identities which relate 
three-point vertices of multipoint correlation functions 
to the single-particle Green function play a crucial role.
These identity relations are derived from the conservation laws for charge and
spin degrees of freedom.

The Ward identity corresponding 
to the charge conservation law has the standard form,
\begin{eqnarray}
\sum_{\mu=0,x,y,z}q_{\mu}\Lambda^c_{\mu}(p+q,p)
=\hat{G}^{-1}(p)-\hat{G}^{-1}(p+q), \label{wicha}
\end{eqnarray}
where $q_{\mu}=(-i\omega,\mbox{\boldmath $q$})$, and 
each components of $\Lambda^c_{\mu}$ are the three-point vertex
functions dressed by electron-electron interaction
for the charge density ($\mu=0$) and the charge current ($\mu=x,y,z$)
In the limit of $q\rightarrow 0$,
the charge current vertex function is given by,
\begin{eqnarray}
\Lambda^c_{\mu}(p,p)=\frac{\partial}{\partial k_{\mu}}
(\varepsilon_k+\Sigma_0+\alpha(\mbox{\boldmath $\sigma$}\cdot
\mbox{\boldmath $\mathcal{L}$}(p))). \label{chac}
\end{eqnarray}

In contrast to Eq.(\ref{wicha}),
the Ward identity for the spin degrees of freedom is affected by
the spin-orbit interaction which breaks the conservation of
the local spin density.
To derive the Ward identity for the spin degrees of freedom,
we start from
the continuity equation for the spin density,
\begin{eqnarray}
\frac{\partial S^{\nu}(x)}{\partial t}+\nabla\cdot
\mbox{\boldmath $J$}^{s\nu}(x)
=\rho_{\rm S}^{\nu}(x).  \label{cons}
\end{eqnarray}
Here, in terms of the spinor field
$\psi^{\dagger}(x)=\frac{1}{\sqrt{V}}
\sum_k c_{k}^{\dagger}\exp(-ikx)$,
%(\psi^{\dagger}_{\uparrow}(x),\psi^{\dagger}_{\downarrow}(x))$ with
%$\psi_{\sigma}^{\dagger}(x)=\frac{1}{\sqrt{V}}
%\sum_k c_{\sigma k}^{\dagger}\exp(-ikx)$,
the $\nu$-component of the spin density operator $S^{\nu}(x)$ ($\nu=x,y,z$) 
and
the corresponding spin current density operator $J^{s\nu}(x)$ are expressed as,
\begin{eqnarray}
S^{\nu}(x)=\psi^{\dagger}(x)\frac{\sigma_{\nu}}{2}
\psi(x),
%S^z(x)=\psi^{\dagger}_{\sigma}(x)\frac{\sigma^z_{\sigma\sigma'}}{2}
%\psi_{\sigma'}(x),
\end{eqnarray} 
\begin{eqnarray}
\mbox{\boldmath $J$}^{s\nu}=\frac{1}{4i}
[\psi^{\dagger}(x)\sigma_{\nu}\nabla\psi(x)
-\{\nabla\psi^{\dagger}(x)\}\sigma_{\nu}\psi(x)]. 
%\mbox{\boldmath J}^{sz}=\frac{\sigma^z_{\sigma\sigma'}}{4i}
%[\psi^{\dagger}_{\sigma}(x)\nabla\psi_{\sigma'}(x)
%-\{\nabla\psi^{\dagger}_{\sigma}(x)\}\psi_{\sigma'}(x)]. 
\end{eqnarray}
In (\ref{cons}),
$\rho_{\rm S}^{\nu}(x)$ is the source term
caused by the spin-orbit interaction 
$\mathcal{H}_{\rm SO}$. From 
the coordinate representation of $\mathcal{H}_{\rm SO}$,
\begin{eqnarray}
{\cal H}_{\rm SO}&=&\alpha\int \frac{dx}{2}[
\psi^{\dagger}(x)
\mbox{\boldmath $\sigma$}\cdot\{\mbox{\boldmath $\mathcal{L}$}_{0}(-i\nabla)
\psi(x)\} \nonumber \\
&-&\{\mbox{\boldmath $\mathcal{L}$}_{0}(-i\nabla)
\psi^{\dagger}(x)\}\cdot\mbox{\boldmath $\sigma$}
\psi(x) 
], \label{fso}
%{\cal H}_{\rm SO}&=&\alpha\int \frac{dx}{2i}\sum_{\sigma,\sigma'}[
%\mbox{\boldmath $\sigma$}_{\sigma\sigma'}
%\cdot\{\mbox{\boldmath $\mathcal{L}$}_{0}(-i\nabla)
%\psi^{\dagger}_{\sigma}(x)\}\psi_{\sigma'}(x) \nonumber \\
%&-&\psi^{\dagger}_{\sigma}(x)
%\mbox{\boldmath $\sigma$}_{\sigma\sigma'}
%\cdot\{\mbox{\boldmath $\mathcal{L}$}_{0}(-i\nabla)
%\psi_{\sigma'}(x)\}], \label{fso}
\end{eqnarray}
we have the explicit form of $\rho_{\rm S}^{\nu}(x)$, 
\begin{eqnarray}
\rho_{\rm S}^{\nu}(x)&=&\frac{\alpha}{2}[\psi^{\dagger}(x)
\mbox{\boldmath $\sigma$}\cdot (\mbox{\boldmath $n$}_{\nu}\times
\mbox{\boldmath {$\mathcal{L}$}}_0(-i\nabla))\psi(x) \nonumber \\
&-&
\{(\mbox{\boldmath $n$}_{\nu}\times
\mbox{\boldmath {$\mathcal{L}$}}_0(-i\nabla))
\cdot\mbox{\boldmath $\sigma$}
\psi^{\dagger}(x)\}\psi(x) \nonumber \\
&-&\psi^{\dagger}(x)i
\mathcal{L}_{0\nu}(-i\nabla)\psi(x)
\nonumber \\
&-&i\{\mathcal{L}_{0\nu}(-i\nabla)\psi^{\dagger}(x)\}
\psi(x)],
%\rho_{\rm S}^z(x)&=&\frac{\alpha}{2}[\psi^{\dagger}_{\sigma}(x)
%\mbox{\boldmath $\sigma$}_{\sigma\sigma'}\cdot ({$\mathcal{L}$}_0(-i\nabla)
%\times \mbox{\boldmath $n$})\psi_{\sigma'}(x) \nonumber \\
%&-&\mbox{\boldmath $\sigma$}_{\sigma\sigma'}
%\cdot \{({$\mathcal{L}$}_0(-i\nabla)\times \mbox{\boldmath $n$})
%\psi^{\dagger}_{\sigma}(x)\}\psi_{\sigma'}(x) \nonumber \\
%&+&\psi^{\dagger}_{\sigma}(x)i
%{$\mathcal{L}$}^z_0(-i\nabla)\psi_{\sigma}(x)
%\nonumber \\
%&+&i\{{$\mathcal{L}$}^z_0(-i\nabla)\psi^{\dagger}_{\sigma}(x)\}
%\psi_{\sigma}(x)], \label{source}
\end{eqnarray} 
where $\mbox{\boldmath $n$}_{\nu}$ is a unit vector along the $\nu$-axis.
In the case of the Rashba spin-orbit interaction which breaks 
the inversion symmetry in the $(0,0,1)$-direction,
$\mbox{\boldmath {$\mathcal{L}$}}_0(k)=
\mbox{\boldmath $t$}_0(k)\times\mbox{\boldmath $n$}_z$.
Here the vector $\mbox{\boldmath $t$}_0(k)=(t_{0x},t_{0y},0)$ 
transforms like $(k_x,k_y,0)$. 
Then, the source term for $S^z$ is of the form,
\begin{eqnarray}
\rho_{\rm S}^z(x)&=&-\frac{\alpha}{2}[\psi^{\dagger}(x)
\mbox{\boldmath $\sigma$}\cdot\mbox{\boldmath $t$}_0(-i\nabla)\psi(x) 
\nonumber \\
&&-\{\mbox{\boldmath $t$}_0(-i\nabla)\psi^{\dagger}(x)\}
\cdot\mbox{\boldmath $\sigma$}\psi(x)],
%\rho_{\rm S}^z(x)&=&\frac{\alpha}{2}[\psi^{\dagger}_{\sigma}(x)
%\mbox{\boldmath $\sigma$}_{\sigma\sigma'}\cdot
%\mbox{\boldmath $t$}_0(-i\nabla)\psi_{\sigma'}(x) 
%\nonumber \\
%&&-\mbox{\boldmath $\sigma$}_{\sigma\sigma'}\cdot
%\{\mbox{\boldmath $t$}_0(-i\nabla)
%\psi^{\dagger}_{\sigma}(x)\}
%\psi_{\sigma'}(x)]
\end{eqnarray}
and for $S^x$,
\begin{eqnarray}
\rho^x_S(x)&=&-\frac{\alpha}{2}[\psi^{\dagger}\sigma_zt_{0x}(-i\nabla)
\psi(x) 
-\{t_{0x}(-i\nabla)\psi^{\dagger}(x)\}\sigma_z\psi(x) \nonumber \\
&+&\psi^{\dagger}(x)i t_{0y}(-i\nabla)\psi(x)
+i\{t_{0y}(-i\nabla)\psi^{\dagger}(x)\}\psi(x)
].
\end{eqnarray}

\begin{figure}
\begin{center}
\includegraphics[width=6cm]{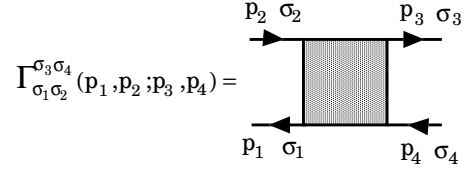}% Here is how to import EPS art
\end{center}
\caption{\label{fig:vertex} Diagrammatic representation of
the four-point vertex with spin indices $\sigma_1$, $\sigma_2$, $\sigma_3$, 
and $\sigma_4$, which is irreducible with respect to
particle-hole pairs.} 
\end{figure}

The standard argument applied to the continuity equation (\ref{cons})
leads the Ward identity for the spin degrees of freedom,\cite{sch}
\begin{eqnarray}
\sum_{\mu=0,x,y,z}q_{\mu}\hat{\Lambda}^{s\nu}_{\mu}(p+q,p)
&=&\frac{\sigma_\nu}{2}\hat{G}^{-1}(p) 
-\hat{G}^{-1}(p+q)\frac{\sigma_\nu}{2}
\nonumber \\
&&+\hat{\mathcal{T}}^{\nu}(p+q,p), \label{swi}
\end{eqnarray}
where $\hat{\Lambda}^{s\nu}_{\mu}(p+q,p)$ ($\mu=0,x,y,z$)
are the three-point vertex functions dressed by electron-electron
interaction
for the $\nu$-component of the spin density ($\mu=0$) and 
the corresponding spin current density ($\mu=x,y,z$). 
$\hat{\mathcal{T}}^{\nu}(p+q,p)$ is the dressed three-point vertex function for
the spin source $\rho_{\rm S}^{\nu}$.
$\hat{\Lambda}^{s\nu}_{\mu}$ and $\hat{\mathcal{T}}^{\nu}$ are 
the $2\times 2$ matrices in the spin space.
The three-point vertex $\hat{\mathcal{T}}^{\nu}(p+q,p)$ is
given by the solution of the following integral equation.
\begin{eqnarray}
&&\hat{\mathcal{T}}^{\nu}(p+\frac{q}{2},p-\frac{q}{2})
=\hat{\mathcal{T}}_0^{\nu}(p+\frac{q}{2},p-\frac{q}{2}) \nonumber \\
&&+\sum_{p'}{\rm tr}[\hat{\Gamma}(p+\frac{q}{2},p-\frac{q}{2};
p'+\frac{q}{2},p'-\frac{q}{2}) \nonumber \\
&&\times\hat{G}(p'+\frac{q}{2})\hat{\mathcal{T}}^{\nu}
(p'+\frac{q}{2},p'-\frac{q}{2})\hat{G}(p'-\frac{q}{2})].\label{inteq}
\end{eqnarray}
Here $\hat{\Gamma}(p_1,p_2;p_3,p_4)$ is the four-point vertex function
which is irreducible with respect to particle-hole pairs.
The diagrammatic representation of $\hat{\Gamma}$ with spin indices
is shown in FIG.\ref{fig:vertex}.
The bare vertex function $\hat{\mathcal{T}}_0^{\nu}$ is defined as
\begin{eqnarray}
&&\hat{\mathcal{T}}_0^{\nu}(p+\frac{q}{2},p-\frac{q}{2})=
\frac{\alpha}{2}
[\mathcal{L}_{0\nu}(\mbox{\boldmath $k$}-\frac{\mbox{\boldmath $q$}}{2})
-\mathcal{L}_{0\nu}(\mbox{\boldmath $k$}+\frac{\mbox{\boldmath $q$}}{2})]
\nonumber \\
&&+\frac{i\alpha}{2}
(\{\mbox{\boldmath $\mathcal{L}$}_0(\mbox{\boldmath $k$}
+\frac{\mbox{\boldmath $q$}}{2})
+\mbox{\boldmath $\mathcal{L}$}_0(\mbox{\boldmath $k$}
-\frac{\mbox{\boldmath $q$}}{2})\}\times
\mbox{\boldmath $\sigma$})\cdot
\mbox{\boldmath $n$}_{\nu}. 
\end{eqnarray}
In the case of the Rashba spin-orbit interaction,
the $z$-component of the bare vertex function is,
\begin{eqnarray}
\hat{\mathcal{T}}_0^z(\mbox{\boldmath $k$}+\frac{\mbox{\boldmath $q$}}{2},
\mbox{\boldmath $k$}-\frac{\mbox{\boldmath $q$}}{2})
=\frac{i\alpha}{2}\mbox{\boldmath $\sigma$}\cdot[
\mbox{\boldmath $t$}_0(\mbox{\boldmath $k$}+\frac{\mbox{\boldmath $q$}}{2})
+\mbox{\boldmath $t$}_0(\mbox{\boldmath $k$}-\frac{\mbox{\boldmath $q$}}{2})],
\end{eqnarray}
and the $x$-component is,
\begin{eqnarray}
&&\hat{\mathcal{T}}_0^x(\mbox{\boldmath $k$}+\frac{\mbox{\boldmath $q$}}{2},
\mbox{\boldmath $k$}-\frac{\mbox{\boldmath $q$}}{2})
=-\frac{\alpha}{2}[t_{0y}(\mbox{\boldmath $k$}
+\frac{\mbox{\boldmath $q$}}{2})-
t_{0y}(\mbox{\boldmath $k$}
-\frac{\mbox{\boldmath $q$}}{2})]    \nonumber \\
&&-\frac{i\alpha\sigma_z}{2}[t_{0x}(\mbox{\boldmath $k$}
+\frac{\mbox{\boldmath $q$}}{2})+
t_{0x}(\mbox{\boldmath $k$}
-\frac{\mbox{\boldmath $q$}}{2})]. \label{tt}
\end{eqnarray} 
The above formulae (\ref{swi})$-$(\ref{tt}) are useful
for the computations of various correlation functions 
considered in the following sections. 

\begin{figure}
\begin{center}
\includegraphics[width=6cm]{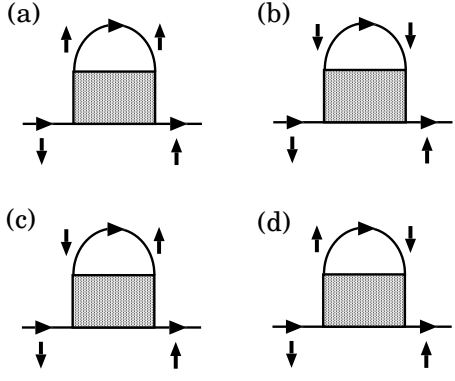}% Here is how to import EPS art
\end{center}
\caption{\label{fig:selfdia} Diagrams representing
the off-diagonal self-energy $\Sigma_{\uparrow\downarrow}$.
The gray box is a four point vertex. The up and down arrows, respectively, 
express the up and down spins.} 
\end{figure}

\begin{figure}
\begin{center}
\includegraphics[width=8cm]{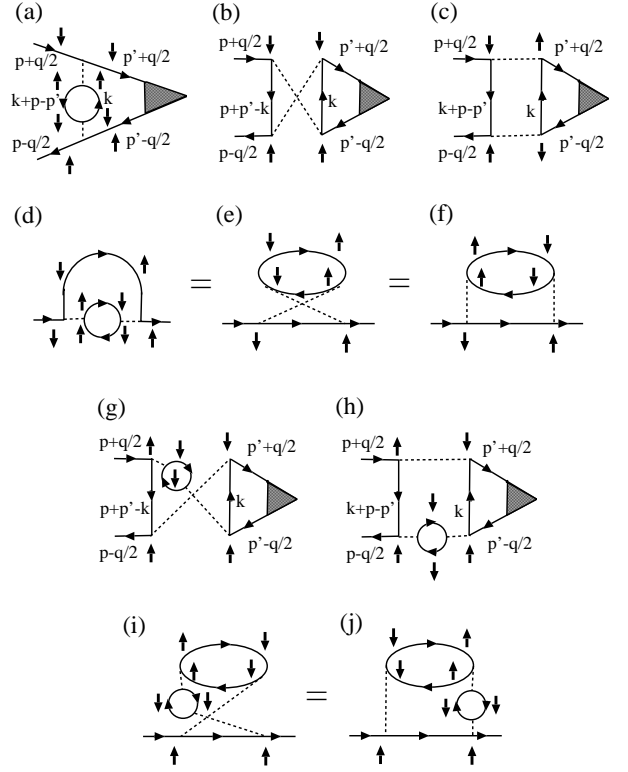}% Here is how to import EPS art
\end{center}
\caption{\label{fig:self-ske} (a), (b), (c) Examples of 
second order diagrams which constitute $\hat{\mathcal{T}}^{\nu}(p+q/2,p-q/2)$.
Broken lines represent
the Coulomb interaction $U$. Black triangle represents 
the bare vertex
$\hat{\mathcal{T}}_0^{\nu}$.
(d), (e) Second-order self-energy diagrams which constitute 
$\sum_{p'}
\Gamma_{\uparrow\downarrow}^{\downarrow\uparrow}(p,p;p',p')
G_{\uparrow\downarrow}(p')$.
The diagrams (d) and (e) are, respectively, related to
the diagrams (a) and (b) via the Ward identity.
(f) A second-order self-energy diagrams which constitute 
$\sum_{p'}
\Gamma_{\uparrow\downarrow}^{\uparrow\downarrow}(p,p;p',p')
G_{\downarrow\uparrow}(p')$.
The diagram (f) is related to the diagram (c) via the Ward identity.
Note that the self-energy
diagrams (d), (e), and (f) are equivalent to each other.
(g), (h)  Examples of 
third-order diagrams for $\hat{\mathcal{T}}^{\nu}(p+q/2,p-q/2)$.
(i) A third-order self-energy diagram
 which constitute
$\sum_{p'}\Gamma_{\uparrow\uparrow}^{\uparrow\downarrow}(p,p;p',p')
G_{\downarrow\uparrow}(p')$. 
The diagram (i) is related to the diagram (g) via the Ward identity.
(j)  A third-order diagram which constitute 
$\sum_{p'}\Gamma_{\uparrow\uparrow}^{\downarrow\uparrow}(p,p;p',p')
G_{\uparrow\downarrow}(p')$. 
The diagram (j) is related to the diagram (h) via the Ward identity.
Note that the diagram (i) is equivalent to the diagram (j).
} 
\end{figure}

In the limit of $q\rightarrow 0$, the Ward identity (\ref{swi})
is reduced to the following relation,
\begin{eqnarray}
\hat{\mathcal{T}}^{\nu}(p,p)=
i\alpha(\mbox{\boldmath $\mathcal{L}$}(p)\times
\mbox{\boldmath $\sigma$})\cdot\mbox{\boldmath $n$}_{\nu}. \label{tl}
\end{eqnarray}
%Here $(\cdot\cdot\cdot)_{\nu}$ indicates the $\nu$-component of 
%the vector $(\cdot\cdot\cdot)$.
The above relation is not so trivial.
Thus it is worth while examining (\ref{tl}) in more details.
Substituting Eq.(\ref{tl}) 
into Eq.(\ref{inteq}) with $q=0$, and 
using the following identity, which is derived from (\ref{gree}),
\begin{eqnarray}
\hat{G}(p)i\alpha (\mbox{\boldmath $\mathcal{L}(p)$}\times
\mbox{\boldmath $\sigma$})_z\hat{G}(p)=\left(
\begin{array}{cc}
0 & G_{\uparrow\downarrow}(p) \\
-G_{\downarrow\uparrow}(p) & 0
\end{array}
\right),
\end{eqnarray}
we find that the identity (\ref{tl}) with $\nu=z$ is equivalent to
the following relations for the self-energy functions,
\begin{eqnarray}
\Sigma_{\uparrow\downarrow}(p)&=&\sum_{p'}[
\Gamma_{\uparrow\downarrow}^{\downarrow\uparrow}(p,p;p',p')
G_{\uparrow\downarrow}(p') \nonumber \\
&&-\Gamma_{\uparrow\downarrow}^{\uparrow\downarrow}(p,p;p',p')
G_{\downarrow\uparrow}(p')], \label{selfid1}
\end{eqnarray}
\begin{eqnarray}
\Sigma_{\downarrow\uparrow}(p)&=&\sum_{p'}[
\Gamma_{\downarrow\uparrow}^{\uparrow\downarrow}(p,p;p',p')
G_{\downarrow\uparrow}(p') \nonumber \\
&-&\Gamma_{\downarrow\uparrow}^{\downarrow\uparrow}(p,p;p',p')
G_{\uparrow\downarrow}(p')], \label{selfid2}
\end{eqnarray}
\begin{eqnarray}
\sum_{p'}\Gamma_{\uparrow\uparrow}^{\downarrow\uparrow}(p,p;p',p')
G_{\uparrow\downarrow}(p')
=\sum_{p'}\Gamma_{\uparrow\uparrow}^{\uparrow\downarrow}(p,p;p',p')
G_{\downarrow\uparrow}(p'), \label{selfid3}
\end{eqnarray}
\begin{eqnarray}
\sum_{p'}\Gamma_{\downarrow\downarrow}^{\downarrow\uparrow}(p,p;p',p')
G_{\uparrow\downarrow}(p')
=\sum_{p'}\Gamma_{\downarrow\downarrow}^{\uparrow\downarrow}(p,p;p',p')
G_{\downarrow\uparrow}(p'). \label{selfid4}
\end{eqnarray}
Here $\Gamma_{\sigma_1\sigma_2}^{\sigma_3\sigma_4}$ 
is four-point vertex functions $\hat{\Gamma}$ which appear in eq.(\ref{inteq}).
The interpretation of the identity relations (\ref{selfid1})-(\ref{selfid4})
in terms of the diagrammatic language
is as follows. 
In FIG.\ref{fig:selfdia}, we depict diagrams which represent
the off-diagonal self-energy $\Sigma_{\uparrow\downarrow}$.
%%%%added
It is noted that the four-point vertex functions $\hat{\Gamma}$ 
in FIG.\ref{fig:selfdia} are obtained by rotating
the four-point vertex shown in FIG.\ref{fig:vertex} by 90 $^{\circ}$.
Thus, the four-point vertex functions
in FIG.\ref{fig:selfdia} are not irreducible
with respect to particle-hole pairs
in the horizontal direction.
%%%%%%%%%%%%%%
Among these diagrams, FIGs.\ref{fig:selfdia}(a) 
and \ref{fig:selfdia}(b) are also expressed 
in the form of FIG.\ref{fig:selfdia}(c),
since in the four-point vertex in FIGs.\ref{fig:selfdia}(a) 
and \ref{fig:selfdia}(b), 
there should be at least one $G_{\uparrow\downarrow}$-line.
We note that the way of expressing $\Sigma_{\uparrow\downarrow}$
in the form of FIG.\ref{fig:selfdia}(c) is not unique. 
The double counting occurs for diagrams which are also 
rewritten in the form of FIG.\ref{fig:selfdia}(d), because 
the four point vertex in FIG.\ref{fig:selfdia}(d) contains two more 
$G_{\uparrow\downarrow}$-lines than that in FIG.\ref{fig:selfdia}(c), 
and thus the diagrams of the form FIG.\ref{fig:selfdia}(d)
are expressed in the form of FIG.\ref{fig:selfdia}(c) in two different ways. 
%%%%%%%%%%added
We show in FIG.\ref{fig:self-ske}
some examples of this double counting
which occurs for $\sum_{p'}
\Gamma_{\uparrow\downarrow}^{\downarrow\uparrow}(p,p;p',p')
G_{\uparrow\downarrow}(p')$ etc.
In FIGs.\ref{fig:self-ske}(a), (b), and (c), we depict some examples of
the second-order diagrams for $\hat{\mathcal{T}}^{\nu}(p+q/2,p-q/2)$.
The Ward identity relates these three different diagrams, 
respectively, to
the diagrams FIGs.\ref{fig:self-ske}(d), (e), and (f), which
express the second-order terms of $\Sigma_{\uparrow\downarrow}$.
The diagrams FIGs.\ref{fig:self-ske}(d) and (e) are parts of
the diagram FIG.\ref{fig:selfdia}(c), 
and the diagrams FIG.\ref{fig:self-ske}(f) is a part of
the diagram FIG.\ref{fig:selfdia}(d).
Apparently, 
these three self-energy diagrams FIGs.\ref{fig:self-ske}(d), (c), and (f)
are equivalent to each other, and double counting of diagrams occurs
in FIG.\ref{fig:selfdia}(c).
%Thus, the correct second-order self-energy diagrams 
%for $\Sigma_{\uparrow\downarrow}$
%are obtained by subtracting the doubly counted diagrams
%from $\sum_{p'}
%\Gamma_{\uparrow\downarrow}^{\downarrow\uparrow}(p,p;p',p')
%G_{\uparrow\downarrow}(p')$.
%%%%%%%%%%%%%%%%%%%%%%%%%%%%%%%%%%%%
Thus, $\Sigma_{\uparrow\downarrow}$ is represented by
FIG.\ref{fig:selfdia}(c) 
from which the duplicate diagrams of the form FIG.\ref{fig:selfdia}(d) are
subtracted. This amounts to the identity (\ref{selfid1}).
We can also give a similar interpretation to Eq.(\ref{selfid2}).
The implication of Eqs.(\ref{selfid3}) and (\ref{selfid4})
is more transparent. 
The left- and right-hand sides of Eq.(\ref{selfid3}) are 
nothing but two equivalent representations
of $\Sigma_{\uparrow\uparrow}$.
%%%%%%%%%%%%%added
In FIG.\ref{fig:self-ske}, we exhibit some examples of this equivalence. 
FIGs.\ref{fig:self-ske}(g) and (h) are
examples of diagrams for $\hat{\mathcal{T}}^{\nu}(p+q/2,p=q/2)$,
which are, respectively, related to the self-energy diagrams
FIGs.\ref{fig:self-ske}(i) and (j), which constitute
$\Sigma_{\uparrow\uparrow}$.
These equivalent self-energy diagrams are, respectively, a part of 
 $\sum_{p'}\Gamma_{\uparrow\uparrow}^{\downarrow\uparrow}(p,p;p',p')
G_{\uparrow\downarrow}(p')$
and a part of 
$\sum_{p'}\Gamma_{\uparrow\uparrow}^{\uparrow\downarrow}(p,p;p',p')
G_{\downarrow\uparrow}(p')$.
This amounts to the relation (\ref{selfid3}).
The relation holds for any higher-order diagrams.
%%%%%%%%%%%%%%%%%%%%%%%%%%%%%%%%%%%%%%%
Eq.(\ref{selfid4}) also expresses $\Sigma_{\downarrow\downarrow}$
in two different manners.

%Thus the relation (\ref{tl}) is interpreted by
%these diagramatic consideration.

\subsection{Magnetism: Pauli and van-Vleck-like spin susceptibilities}

As mentioned in the introduction,
one of the unique magnetic properties in inversion-symmetry-broken
systems is the existence of a van-Vleck-like spin susceptibility
which stems from magnetic response of electrons occupying
the momentum space sandwiched 
between spin-orbit splitted two Fermi surfaces. 
In this section we obtain general formulae for
the Pauli and van-Vleck-like spin susceptibilities taking into account
electron correlation effects in a formally exact way.
The uniform spin susceptibility is easily obtained from (\ref{gree}) and 
(\ref{grepm})
by using
\begin{eqnarray}
\chi_{\mu\mu}=-\frac{d}{d H_{\mu}}T\sum_{\varepsilon_n}
\sum_k\mu_{\rm B}{\rm tr}[\sigma_{\mu}\hat{G}(p)]\vert_{H_{\mu}\rightarrow 0}.
\label{chi}
\end{eqnarray} 

In the case of cubic systems, 
a straightforward calculation gives,
\begin{eqnarray}
\chi_{zz}&=&\mu_{\rm B}^2\sum_{\tau=\pm}\sum_k\frac{z_{k\tau}}{4T
\cosh^2(\frac{\varepsilon^{*}_{k\tau}}{2T})}
\Lambda_{\rm P}^{\rm cub}(\varepsilon_{k\tau}^{*},\mbox{\boldmath $k$}) 
\nonumber \\
&+&\mu_{\rm B}^2\sum_{\tau=\pm}\sum_k\frac{
-\tau f(\varepsilon^{*}_{k\tau})
%\tau\tanh(\frac{\varepsilon^{*}_{k\tau}}{2T}))
}{\alpha |{\rm Re}~
\mbox{\boldmath $\mathcal{L}$}
(\varepsilon_{k\tau}^{*},\mbox{\boldmath $k$})|}
\Lambda^{\rm cub}_{\rm V}
(\varepsilon_{k\tau}^{*},\mbox{\boldmath $k$}),
\label{chicu}
\end{eqnarray}
with $f(\varepsilon)$ the Fermi distribution function.
Here the three-point vertex functions $\Lambda^{\rm cub}_{\rm P}$ 
and $\Lambda^{\rm cub}_{\rm V}$ are,
\begin{eqnarray}
\Lambda^{\rm cub}_{\rm P}(p)&=&\hat{\mathcal{L}}_{0z}^2
(1-\frac{1}{\mu_{\rm B}}\frac{\partial \Sigma_0(p)}{\partial H_z})
\nonumber \\
&-&\frac{\hat{\mathcal{L}}_{0z}
\hat{\mathcal{L}}_x}{\mu_{\rm B}}
\frac{\partial \Sigma_x(p)}{\partial H_z}
-\frac{\hat{\mathcal{L}}_{0z}
\hat{\mathcal{L}}_y}{\mu_{\rm B}}
\frac{\partial \Sigma_y(p)}{\partial H_z},\label{tv1}
\end{eqnarray}
\begin{eqnarray}
\Lambda^{\rm cub}_{\rm V}(p)&=&(\hat{\mathcal{L}}_x^2
+\hat{\mathcal{L}}_y^2)
(1-\frac{1}{\mu_{\rm B}}\frac{\partial \Sigma_0(p)}{\partial H_z})
\nonumber \\
&+&\frac{\hat{\mathcal{L}}_{0z}
\hat{\mathcal{L}}_x}{\mu_{\rm B}}
\frac{\partial \Sigma_x(p)}{\partial H_z}
+\frac{\hat{\mathcal{L}}_{0z}
\hat{\mathcal{L}}_y}{\mu_{\rm B}}
\frac{\partial \Sigma_y(p)}{\partial H_z}.\label{tv2}
\end{eqnarray}
%Here we have used $\hat{\mbox{\boldmath $\mathcal{L}$}}_z=
%\hat{\mbox{\boldmath $\mathcal{L}$}}_{0z}$ for $H_z=0$.
The first and second terms of (\ref{chicu}) are, respectively, 
the Pauli and van-Vleck-like terms.
The van-Vleck-like term is not much affected by superconducting transition
if the magnitude of the spin-orbit splitting 
$\alpha |\mbox{\boldmath $\mathcal{L}$}(p)|$ is sufficiently larger than
the superconducting gap.

In the case of the tetragonal symmetry with a potential gradient 
along $\mbox{\boldmath $n$}=(0,0,1)$ 
(i.e. the Rashba interaction),
$\mbox{\boldmath {$\mathcal{L}$}}_0(k)$ is parametrized as   
$\mbox{\boldmath $t$}_0(k)\times\mbox{\boldmath $n$}$.
The vector $(t_{0x},t_{0y},0)$ transforms like $(k_x,k_y,0)$.
Then, using a vector 
$\mbox{\boldmath $t$}(p)=\mbox{\boldmath $n$}\times
\mbox{\boldmath {$\mathcal{L}$}}(p)$ and
a unit vector $\hat{\mbox{\boldmath $t$}}(p)=
\mbox{\boldmath $t$}(p)/|\mbox{\boldmath $t$}(p)|$,
we express the spin susceptibility as, 
\begin{eqnarray}
\chi_{zz}=\mu_{\rm B}^2\sum_{\tau=\pm}\sum_k\frac{
-\tau f(\varepsilon^{*}_{k\tau})
%\tau\tanh(\frac{\varepsilon^{*}_{k\tau}}{2T})
}{\alpha 
|{\rm Re}~\mbox{\boldmath $t$}(\varepsilon_{k\tau}^{*},\mbox{\boldmath $k$})|}
\Lambda^{sz}(\varepsilon_{k\tau}^{*},\mbox{\boldmath $k$}), \label{chizz}
\end{eqnarray}
\begin{eqnarray}
\chi_{xx}=\chi_{xx}^{\rm Pauli}+\chi_{xx}^{\rm VV}, \label{chixx}
\end{eqnarray}
\begin{eqnarray}
\chi_{xx}^{\rm Pauli}=\mu_{\rm B}^2\sum_{\tau=\pm}\sum_k\frac{z_{k\tau}}{4T
\cosh^2(\frac{\varepsilon^{*}_{k\tau}}{2T})}
\hat{t}_y\Lambda_{\tau}^{sx}(\varepsilon_{k\tau}^{*},\mbox{\boldmath $k$}),
\label{chixp} 
\end{eqnarray}
\begin{eqnarray}
\chi_{xx}^{\rm VV}=\mu_{\rm B}^2\sum_{\tau=\pm}\sum_k\frac{
-\tau f(\varepsilon^{*}_{k\tau})
%\tau\tanh(\frac{\varepsilon^{*}_{k\tau}}{2T})
}{\alpha 
|{\rm Re}~\mbox{\boldmath $t$}(\varepsilon_{k\tau}^{*},\mbox{\boldmath $k$})|}
\hat{t}_x\Lambda^{sx}_{+-}(\varepsilon_{k\tau}^{*},\mbox{\boldmath $k$}).
\label{chixv}
\end{eqnarray}
The three-point vertices $\Lambda^{sz}$, $\Lambda^{sx}_{+-}$, and 
$\Lambda^{sx}_{\tau}$ in the above expressions are given by
\begin{eqnarray}
\Lambda^{sz}(p)=
1-\frac{1}{\mu_{\rm B}}\frac{\partial \Sigma_0(p)}{\partial H_z},
\label{svert1}
\end{eqnarray}
\begin{eqnarray}
\Lambda^{sx}_{\tau}(p)=\hat{t}_y
(1-\frac{1}{\mu_{\rm B}}\frac{\partial 
\Sigma_{x}}{\partial H_x})+\frac{\hat{t}_x}
{\mu_{\rm B}}\frac{\partial 
\Sigma_{y}}{\partial H_x}
-\frac{\tau}
{\mu_{\rm B}}\frac{\partial \Sigma_0}{\partial H_x},
\label{svert2}
\end{eqnarray}
\begin{eqnarray}
\Lambda_{+-}^{sx}(p)=\hat{t}_x(1-\frac{1}{\mu_{\rm B}}\frac{\partial 
\Sigma_{x}}{\partial H_x})
-\frac{\hat{t}_y}{\mu_{\rm B}}
\frac{\partial 
\Sigma_{y}}{\partial H_x}.
\label{svert3}
\end{eqnarray}
As seen from (\ref{chizz}) and (\ref{chixx}), 
$\chi_{zz}$ is given only by the van-Vleck-like susceptibility, 
while $\chi_{xx}$ consists
of both the Pauli and van-Vleck-like terms.

We note that in the limit of $\alpha\rightarrow 0$ (the inversion symmetry 
and the spin rotation symmetry
are recovered), 
\begin{eqnarray}
\frac{\partial \Sigma_{x}}{\partial H_x}
\rightarrow\frac{\partial \Sigma_0}{\partial H_z}.
\end{eqnarray} 
%%%%%%%% added %%%%%%%%%%%%%%%%%%%%%
%because the rotation in the spin space $S_x\rightarrow S_z$ transforms
%$\Sigma_x=(\Sigma_{\uparrow\downarrow}+\Sigma_{\downarrow\uparrow})/2$
%into $\Sigma_{\uparrow\uparrow}=\Sigma_{\downarrow\downarrow}$.
%%%%%%%%%%%%%%%%%%%%%%%%added%%%%
This relation is easily verified by rotating the spin axes as 
$\sigma_x\rightarrow \sigma_z$, $\sigma_z\rightarrow -\sigma_x$, and
$\sigma_y\rightarrow\sigma_y$, and noticing that this rotation
transforms $(\Sigma_{\uparrow\downarrow}+\Sigma_{\downarrow\uparrow})/2$ into 
$(\Sigma_{\uparrow\uparrow}-\Sigma_{\downarrow\downarrow})/2$.
%%%%%%%%%%%%%%%%%%%%%%%%%%%%%%%%%%%%%
Thus, Eqs.(\ref{chizz}) and (\ref{chixx}) leads $\chi_{zz}=\chi_{xx}$
in the limit of $\alpha\rightarrow 0$.

We see from (\ref{chizz}) and (\ref{chixx}) that both the Pauli term and
the van-Vleck-like term are enhanced by the factors $\Lambda^{sz}$,
$\Lambda^{sz}_{+-}$, and $\Lambda^{sz}_{\tau}$.
If the density of states weakly depends on energy,
and the spin-orbit splitting is much smaller than 
the Fermi energy, these three enhancement factors  
take values of the same order.
However, when the energy dependence of the density of states
is substantial, the enhancement of the van-Vleck-like 
susceptibility due to the electron correlation effect 
may be quite distinct from
that of the Pauli term.
We will consider such an example in Sec.III in connection with
the heavy fermion superconductor CePt$_3$Si.

For experimental studies on superconducting states, 
it is important to discriminate between
the Pauli susceptibility which decreases in the spin-singlet
superconducting state, and the orbital susceptibility
which is not affected by the superconducting transition.
Usually, this discrimination 
is achieved by experimentally observing the temperature dependence
of the spin susceptibility in the normal state, and identifying
the temperature-independent part with the orbital term.
However, this idea fails for noncentrosymmetric superconductors
because of the existence of the van-Vleck-like susceptibility 
$\chi_{xx}^{\rm VV}$.
When the spin-orbit splitting is much larger than 
the superconducting gap as in the case of CePt$_3$Si, CeRhSi$_3$, and
CeIrSi$_3$,
the van-Vleck-like term $\chi_{xx}^{\rm VV}$ is not much influenced by
the superconducting transition and takes a finite value even at $T=0$,
though it exhibits temperature dependence at high temperatures
like the Pauli susceptibility.

\begin{figure}
\begin{center}
\includegraphics[width=8cm]{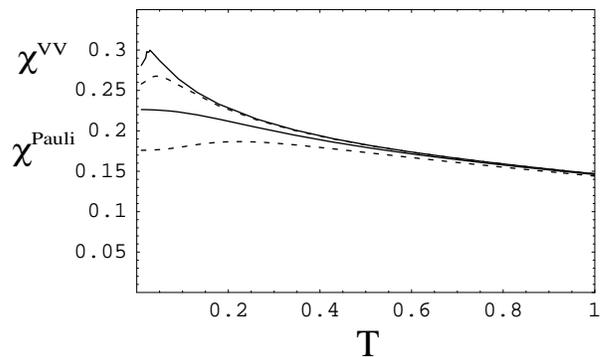}% Here is how to import EPS art
\end{center}
\caption{\label{fig:spinsus} $\chi_{xx}^{\rm Pauli}$ and $\chi_{xx}^{\rm VV}$
versus $T$.  Each lines are, respectively, 
$\chi_{xx}^{\rm VV}$ for $\alpha=0.1$ (top solid line), 
$\chi_{xx}^{\rm Pauli}$ for $\alpha=0.1$ (top dotted line),
$\chi_{xx}^{\rm VV}$ for $\alpha=0.5$ (bottom solid line), 
and $\chi_{xx}^{\rm Pauli}$ for $\alpha=0.5$ (bottom dotted line).} 
\end{figure}

Here we demonstrate to what extent $\chi_{xx}^{\rm VV}$ varies
as a function of temperature by using a simple model.
We consider a two-dimensional electron system
on a square lattice with the energy momentum dispersion relation,
\begin{eqnarray}
\varepsilon_k=-2(\cos k_x+\cos k_y), \label{sq}
\end{eqnarray}
and the Rashba spin-orbit interaction $\alpha (\mbox{\boldmath $t$}_{0}(k)
\times \mbox{\boldmath $n$})\cdot\mbox{\boldmath $\sigma$}$.
We put the hopping integral $t=1$ in (\ref{sq}).
To preserve the periodicity of the energy band in the momentum space,
we assume the form of $\mbox{\boldmath $t$}_{0}(k)$ as 
\begin{eqnarray}
\mbox{\boldmath $t$}_{0}(k)=
(\sin\frac{k_x}{2},\sin\frac{k_y}{2},0). \label{sqra}
\end{eqnarray}
This model can not be directly related with heavy fermion
superconductors such as CePt$_3$Si, CeRhSi$_3$, etc. which have
the complicated three-dimensional Fermi surfaces.
Nevertheless, this simple model is quite instructive in that
it demonstrates the strong temperature dependence of $\chi_{xx}^{\rm VV}$.
For simplicity we neglect electron-electron interaction, and consider
the half-filling case $n=1$, in which
the van Hove singularity of the density of states gives rise to
the temperature-dependence of the spin susceptibility.
For $\alpha\neq 0$, 
to keep the electron density $n=1$, we need to add 
a counter term
\begin{eqnarray}
-\delta\mu \sum_k c_{k}^{\dagger}c_k,
\end{eqnarray}
with $\delta\mu=-\alpha^2/8$ to the Hamiltonian.
%Then the Pauli and van-Vleck-like susceptibilities for an
%inplane magnetic field, $\chi_{xx}^{\rm Pauli}$ and $\chi_{xx}^{\rm VV}$,
%are easily calculated.
The calculated results of $\chi_{xx}^{\rm Pauli}$ and $\chi_{xx}^{\rm VV}$ 
for some values of $\alpha$ are plotted as a function of temperature in 
FIG.\ref{fig:spinsus}.
It is seen that as $T$ increases, the curves of $\chi_{xx}^{\rm VV}$ 
and $\chi_{xx}^{\rm Pauli}$ approach to each other, 
and for $T\gg\alpha$, there is no substantial difference between them. 
The van-Vleck-like term $\chi_{xx}^{\rm VV}$ varies as
a function of temperature, reflecting
the singular energy dependence of the density of states in the model (\ref{sq}).
If we take into account electron correlation effects, 
the energy dependence of the density of states is much enhanced,
and thus, as $T$ is changed, $\chi_{xx}^{\rm VV}$ varies 
more drastically than the non-interacting case.
This calculation suggests that 
%in contrast to
%usual systems with inversion center, 
it is almost impossible to
distinguish the van-Vleck-like term from the Pauli susceptibility
merely by analyzing experimental data of temperature dependence of
the spin susceptibility in the normal state.
%This fact is crucial for the experimental determination
%of the symmetry of the Cooper pair.

\begin{figure}
\begin{center}
\includegraphics[width=8cm]{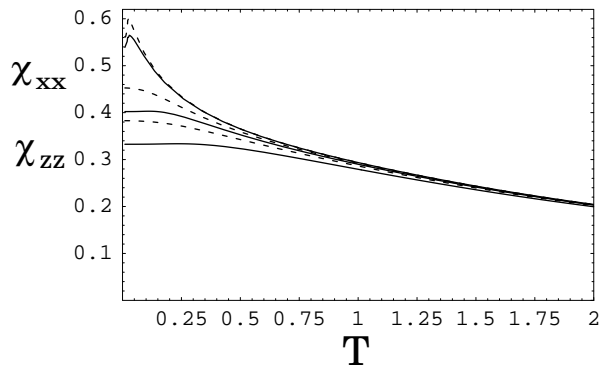}% Here is how to import EPS art
\end{center}
\caption{\label{fig:chiani} $\chi_{xx}$ and $\chi_{zz}$
versus $T$. Each lines are, respectively, 
$\chi_{zz}$ for $\alpha=0.1$ (top dotted line), 
$\chi_{xx}$ for $\alpha=0.1$ (top solid line),
$\chi_{zz}$ for $\alpha=0.5$ (middle dotted line),
$\chi_{xx}$ for $\alpha=0.5$ (middle solid line),
$\chi_{zz}$ for $\alpha=1.0$ (bottom dotted line), and
$\chi_{xx}$ for $\alpha=1.0$ (bottom solid line).} 
\end{figure}

It is also worth while examining the anisotropy
of the spin susceptibility due to the spin-orbit interaction. 
The calculated results of $\chi_{xx}$ and $\chi_{zz}$ as
a function of $T$ are plotted in FIG.\ref{fig:chiani} 
for several values of $\alpha$.
It shows that the anisotropy becomes large for 
$T<\alpha |\mbox{\boldmath $t$}_{0k}|$.
The magnitude of the anisotropy $|\chi_{xx}-\chi_{zz}|/\chi_{xx}$
is at most of order $\sim \alpha |\mbox{\boldmath $t$}_{0k}|/E_F$,
as easily expected from (\ref{chizz}) and (\ref{chixx}).
According to the LDA calculations for the heavy fermion superconductor
CePt$_3$Si, the ratio $\alpha |\mbox{\boldmath $t$}_{0k}|/E_F$
is of order $\sim 0.1$.\cite{hari,sam}
Although the de Haas van Alphen experiment 
for this system has not yet detected
the spin-orbit splitted Fermi surfaces,
the measurement for a related system LaPt$_3$Si support
the existence of the spin-orbit splitting of which the magnitude is
of this order.\cite{hari}
Thus, it is expected that the anisotropy of the spin susceptibility
for CePt$_3$Si is at most $\sim 0.1$, which seems to be consistent
with the experimental observation of the susceptibility.\cite{take,met}

\subsection{Transport coefficients}

Generally, the absence of inversion symmetry
brings about remarkable transport properties such as
the anomalous Hall effect,\cite{kl,jun} the thermal 
anomalous Hall effect, the spin Hall effect,\cite{she,she2}
and magnetoelectric effects.\cite{lev,ede0}
In this section,
we argue these phenomena with particular emphasis on
the role of electron correlation effects.
We mainly consider the case of 
the Rashba spin-orbit interaction 
with $\mbox{\boldmath $\mathcal{L}$}_0=\mbox{\boldmath $t$}_0\times
\mbox{\boldmath $n$}$ in which the above-mentioned effects are 
more salient than the case of cubic systems with the Dresselhaus interaction.

\subsubsection{Anomalous Hall effect}

Here, we consider the anomalous Hall effect due to the
Rashba spin-orbit interaction.
The general theoretical description of the 
anomalous Hall effect due to the spin-orbit interaction 
was given by Karplus-Luttinger many years ago.\cite{kl}
The key ingredient of the Karplus-Luttinger-type anomalous Hall effect
is the existence of a dissipationless transverse current which is
induced by the paramagnetic Zeeman effect and stems
from the anomalous velocity, $\nabla_k \alpha (\mbox{\boldmath $t$}_0(k)
\times\mbox{\boldmath $n$})\cdot\mbox{\boldmath $\sigma$}$.
In the case of the Rashba spin-orbit interaction,
there is no $z$-component of the anomalous velocity, and thus  
this effect exists only for an applied magnetic field along 
the $z$-direction.
In heavy fermion systems, the spin-orbit interaction
caused by heavy atoms like Ce and U 
may also give rise to the anomalous Hall effect
as experimentally observed in several systems such as
CeRu$_2$Si$_2$, CeAl$_3$, and UPt$_3$.\cite{anh,ahe1,ahe2,yam1,yam2}
The strong anisotropic dependence on an external magnetic field
of the Rashba-interaction-induced anomalous Hall effect makes a clear
difference from the conventional anomalous Hall effect observed
in the systems with inversion symmetry.
Note that in the case of cubic systems with the Dresselhaus interaction,
such a strong anisotropy does not exist.

The Hall conductivity for a magnetic field in the $z$-direction
is defined as
\begin{eqnarray}
\sigma_{xy}=\lim_{\omega\rightarrow 0}
\frac{1}{i\omega}K_{xy}(i\omega_n)|_{i\omega_n\rightarrow \omega+i0},
\end{eqnarray}
\begin{eqnarray}
K_{xy}(i\omega_n)=
\int^{1/T}_0 d\tau \langle T_{\tau}\{J_x(\tau)J_y(0)\}
\rangle e^{i\omega_n \tau}.
\end{eqnarray}
The total charge current $J_{\mu}$ is,
\begin{eqnarray}
J_{\mu}=e\sum_kc^{\dagger}_k\hat{v}_{k\mu}c_k,  
\end{eqnarray}
where the velocity $\hat{v}_{k\mu}$ is
\begin{eqnarray}
\hat{v}_{k\mu}=\nabla_{k_\mu}(\varepsilon_k+\alpha
\mbox{\boldmath $\sigma$}\cdot(\mbox{\boldmath $t$}_0(\mbox{\boldmath $k$})
\times\mbox{\boldmath $n$})). \label{cv0}
\end{eqnarray}

Since the anomalous Hall effect is caused by the interplay between 
the spin-orbit interaction and the Zeeman coupling
with an external magnetic field, we neglect the orbital diamagnetic
effect of the magnetic field which yields the normal Hall effect.
Then the Hamiltonian under consideration is given by Eqs.(\ref{ham})
and (\ref{zee}) with $\mbox{\boldmath $H$}=(0,0,H_z)$. From Eq.(\ref{chac}),
the charge velocity vertex fully dressed by electron-electron interaction
is given by,
\begin{eqnarray}
\hat{\Lambda}^c_{\mu}(p,p)&=&\nabla_{k_{\mu}}(\varepsilon_k+\Sigma_0(p))
+\nabla_{k_{\mu}}\Sigma_z(p)\sigma^z \nonumber \\
&&+\alpha\nabla_{k_{\mu}}(t_y(p)\sigma^x-t_x(p)\sigma^y).\label{cv}
\end{eqnarray} 
The last term of the right-hand side of (\ref{cv}) is the anomalous velocity.
In the above expression of $\hat{\Lambda}^c_{\mu}$, the vertex corrections
associated with dissipative processes which are
important for the total momentum conservation\cite{yy} are not included.
The dissipative vertex corrections cannot be taken into account by
the procedure presented above. 
However, their contributions to the anomalous Hall
conductivity are negligible 
in the case that the spin-orbit-splitting is much larger than
the quasiparticle damping because of the following reason.
According to the general argument of the Fermi liquid theory
for non-equilibrium transport coefficients,
the dissipative vertex corrections are obtained by
extracting the particle-hole pair $G^{R}G^{A}$ with
$G^{R(A)}$ the retarded (advanced) single-particle Green function
after the analytic continuation of $K_{xy}(i\omega_n)$,
since only $G^{R}G^{A}$ is singular in the limit of $\omega\rightarrow 0$
for infinitesimally small quasiparticle damping.\cite{eli}
In our case, as will be shown below, the particle-hole pair
$G_{+}G_{+}$ and $G_{-}G_{-}$ do not enter into the expression
of the anomalous Hall conductivity, and only the pair $G_{+}G_{-}$
contributes.
When the spin-orbit-splitting is 
sufficiently larger than the quasiparticle damping, 
the particle-hole pairs $G_{+}^{R}G_{-}^{A}$ and $G_{+}^{R}G_{-}^{A}$
are not singular, and do not yield dissipative processes.
Thus, we can ignore the dissipative vertex corrections
for systems with large spin-orbit splitting.
As a matter of fact, for the noncentrosymmetric
heavy fermion superconductors CePt$_3$Si and CeRhSi$_3$, 
the typical size of 
the spin-orbit-splitting is of order $0.1 E_F$ with $E_F$ 
the Fermi energy, and dominates over 
the quasiparticle damping.\cite{hari,sam,kim2}
Then,
the current-current correlation function $K_{xy}$ is expressed in terms of
the Green functions,
\begin{eqnarray}
K_{xy}(i\omega)&=&T\sum_{\varepsilon_n}
\sum_k{\rm tr}[\hat{\Lambda}^c_x(p+q,p)\hat{G}(p+q) \nonumber \\
&&\times \hat{v}_{ky}
%\alpha\nabla_{k_{y}}(t_{0y}(\mbox{\boldmath $k$})\sigma^x-t_{0x}
%(\mbox{\boldmath $k$})\sigma^y)
%\hat{\Lambda}^c_y(p,p+q)
\hat{G}(p)]. \label{kxy}
\end{eqnarray}
Here $q=(i\omega,0,0,0)$. 
Collecting terms linear in $H_z$ for $H_z\rightarrow 0$,
we find that only the anomalous velocity term
in Eqs.(\ref{cv0}) and (\ref{cv}) gives non-vanishing contributions to 
$K_{xy}(i\omega)$, which have the form 
$\sum_pf(\mbox{\boldmath $k$}){\rm tr}
[\sigma^x\hat{G}(p)\sigma^y\hat{G}(p+q)]$ 
or $\sum_pf(\mbox{\boldmath $k$}){\rm tr}
[\sigma^y\hat{G}(p)\sigma^x\hat{G}(p+q)]$, where
$f(\mbox{\boldmath $k$})$ is the even function of momentum 
$\mbox{\boldmath $k$}$.
Using the assumption that
the quasiparticle damping is much smaller than the spin-orbit-splitting,
we obtain,
\begin{eqnarray}
&&\frac{d}{dH_z}
\sum_pf(\mbox{\boldmath $k$}){\rm tr}[\sigma^y\hat{G}(p)\sigma^x\hat{G}(p+q)]
|_{H_z\rightarrow 0}
=  \nonumber \\
&&i\sum_p[\frac{d}{dH_z}(\xi_{+}^2(p)\xi_{+}^2(p+q)-
\xi_{-}^2(p)\xi_{-}^2(p+q))|_{H_z\rightarrow 0} \nonumber \\
&&\times(G_{+}(p)G_{-}(p+q)-G_{-}(p)G_{+}(p+q)) \nonumber \\
&&+\frac{d}{dH_z}(\xi_{+}^2(p)\xi_{-}^2(p+q)-
\xi_{-}^2(p)\xi_{+}^2(p+q))|_{H_z\rightarrow 0} \nonumber \\
&&\times(G_{+}(p)G_{+}(p+q)
-G_{-}(p)G_{-}(p+q))]. \label{dhg}
\end{eqnarray}
The last term of the right-hand side of (\ref{dhg}) does not
contribute to ${\rm Re}~\sigma_{xy}$.
In the limit of $\omega\rightarrow 0$,
\begin{eqnarray}
&&\frac{d}{dH_z}(\xi_{+}^2(p)\xi_{+}^2(p+q)-
\xi_{-}^2(p)\xi_{-}^2(p+q))|_{H_z\rightarrow 0}  \nonumber \\
&&=-\frac{1}{\alpha |{\rm Re}~\mbox{\boldmath $t$}(p)|}(\mu_{\rm B}-
\frac{\partial \Sigma_0(p)}{\partial H_z}). \label{dh}
\end{eqnarray}
Using (\ref{kxy}), (\ref{dhg}), and (\ref{dh}), we end up with,
\begin{eqnarray}
\frac{{\rm Re}~\sigma_{xy}^{\rm AHE}}{H_z}&=&
e^2\mu_{\rm B}\sum_{\tau=\pm}\sum_k
%\tau\tanh(\frac{\varepsilon^{*}_{k\tau}}{2T})
\frac{-\tau f(\varepsilon^{*}_{k\tau})
%[\Lambda^{sx}_{+-}
%(k,\varepsilon_{k\tau}^{*})]^2
\Lambda^{sz}(
\varepsilon_{k\tau}^{*},\mbox{\boldmath $k$})}
{2\alpha|{\rm Re}~\mbox{\boldmath $t$}
(\varepsilon_{k\tau}^{*},\mbox{\boldmath $k$})|^3} \nonumber \\
&&\times (\partial_{k_{x}}t_{x\tau}\partial_{k_{y}}t_{0y}
-\partial_{k_{x}}t_{y\tau}\partial_{k_{y}}t_{0x}). \label{ahe}
\end{eqnarray}
Here $\partial_{k_{\mu}}t_{\nu\tau}\equiv\partial_{k_{\mu}}t_{\nu}
(\varepsilon,\mbox{\boldmath $k$})|_{\varepsilon=\varepsilon_{k\tau}^{*}}$;
i.e. $\partial_{k_{\mu}}$ does not operate on 
$\varepsilon_{k\tau}^{*}$ in the argument of $t_{\nu}(p)$.
Since we have postulated that the spin-orbit splitting is much larger than the
quasiparticle damping, the anomalous Hall conductivity
$\sigma^{\rm AHE}_{xy}$ is not involved with any relaxation time,
and thus determined only by dissipationless processes.
It is noted that the anomalous Hall conductivity is enhanced by
the factor $\Lambda^{sz}$ which is equivalent to the
enhancement factor of $\chi_{zz}$, Eq.(\ref{chizz}).
In the heavy fermion system CePt$_3$Si, 
this enhancement factor is of order $\sim 60$, and
the detection of the anomalous Hall effect is feasible in
such strongly correlated electron systems. 
We would like to stress that in the expression of
the anomalous Hall conductivity (\ref{ahe}), not only
electrons in the vicinity of the Fermi surface but also 
all electrons in the region of the Brillouin zone sandwiched between
the spin-orbit-splitted two Fermi surfaces contribute, in
accordance with the fact that $\chi_{zz}$ is dominated by 
the van-Vleck-like susceptibility.

\subsubsection{Thermal anomalous Hall effect}

We now consider the thermal anomalous Hall effect, which
is the anomalous Hall effect for the heat current.
To simplify the following analysis, we assume that the energy current
due to the interaction between quasiparticles is negligible, and thus
the heat current is mainly
carried by nearly independent quasiparticles.
We would like to discuss
the validity of this assumption in the end of this section.
Then, we can obtain 
the thermal anomalous Hall conductivity by using a method similar to 
that used in the previous section.
Using the heat current related to the single-particle energy, 
\begin{eqnarray}
J_{Q\mu}=\sum_kc^{\dagger}_k\frac{1}{2}[\hat{v}_{k\mu}\hat{H}(p)+
\hat{H(p)}\hat{v}_{k\mu}]c_k,  
\end{eqnarray}
we define the Hall conductivity for the heat current as,
\begin{eqnarray}
\kappa_{xy}=\frac{1}{T}(L^{(2)}_{xy}
-\sum_{\mu\nu}L^{(1)}_{x\mu}L^{(0)-1}_{\mu\nu}L^{(1)}_{\nu y}), \label{hh}
%\kappa_{xy}=\frac{1}{T}(L^{(2)}_{xy}
%-\frac{(L^{(1)}_{xy})^2}{L^{(0)}_{xy}}), \label{hh}
\end{eqnarray}
where $L^{(0)}_{\mu\nu}$ is equal to the conductivity tensor
$\sigma_{\mu\nu}$, and,
%where $L^{(0)}_{xy}$ is equal to the Hall conductivity $\sigma_{xy}$, and,
\begin{eqnarray}
L^{(m)}_{\mu\nu}=\lim_{\omega\rightarrow 0}\frac{1}{i\omega}
K_{\mu\nu}^{(m)}(i\omega_n)|_{i\omega\rightarrow \omega+i0},
\quad m=1,2,
\end{eqnarray}
%\begin{eqnarray}
%L^{(2)}_{xy}=\lim_{\omega\rightarrow 0}\frac{1}{i\omega}
%K_{xy}^{(2)}(i\omega_n)|_{i\omega\rightarrow \omega+i0},
%\end{eqnarray}
\begin{eqnarray}
K_{\mu\nu}^{(1)}(i\omega_n)=
\int^{1/T}_0 d\tau \langle T_{\tau}\{J_{Q\mu}(\tau)J_{\nu}(0)\}
\rangle e^{i\omega_n \tau},
\end{eqnarray}
\begin{eqnarray}
K^{(2)}_{\mu\nu}(i\omega_n)=
\int^{1/T}_0 d\tau \langle T_{\tau}\{J_{Q\mu}(\tau)J_{Q\nu}(0)\}
\rangle e^{i\omega_n \tau}.
\end{eqnarray}
%We neglect the second term of Eq.(\ref{hh}) for a while.
%If the band structure of the system is particle-hole symmetric, 
%this assumption is valid.  
Extending the argument in the previous
subsection straightforwardly to the present case,
we find that the anomalous Hall effect gives rise to the following
terms to $L_{xy}^{(1)}$ and $L_{xy}^{(2)}$,
\begin{eqnarray}
&&\frac{L_{xy}^{(m) {\rm AHE}}}{H_z}=e^{2-m}
\mu_{\rm B}\sum_{\tau=\pm}\sum_k(-\tau)
(\varepsilon_{k\tau}^{*})^m f(\varepsilon^{*}_{k\tau})
%\tanh(\frac{\varepsilon^{*}_{k\tau}}{2T}) 
\nonumber \\
&&\times
\frac{
%[\Lambda^{sx}_{+-}
%(k,\varepsilon_{k\tau}^{*})]^2
\Lambda^{sz}(\varepsilon_{k\tau}^{*},
\mbox{\boldmath $k$})}
{2\alpha|{\rm Re}~\mbox{\boldmath $t$}
(\varepsilon_{k\tau}^{*},\mbox{\boldmath $k$})|^3} 
(\partial_{k_{x}}t_{x\tau}\partial_{k_{y}}t_{0y}
-\partial_{k_{x}}t_{y\tau}\partial_{k_{y}}t_{0x}), \label{tahll}
\end{eqnarray}
with $m=1,2$.
Then, the expression for the thermal anomalous Hall conductivity 
$\kappa_{xy}^{\rm AHE}$ is given by
Eqs.(\ref{hh}) and (\ref{tahll}).
%\begin{eqnarray}
%&&\frac{{\rm Re}~\kappa_{xy}^{\rm AHE}}{H_z}=
%\frac{\mu_{\rm B}}{T}\sum_{\tau=\pm}\sum_k\tau
%(\varepsilon_{k\tau}^{*})^2\tanh(
%\frac{\varepsilon^{*}_{k\tau}}{2T}) \nonumber \\
%&&\times
%\frac{
%%[\Lambda^{sx}_{+-}
%(k,\varepsilon_{k\tau}^{*})]^2
%\Lambda^z(\varepsilon_{k\tau}^{*},
%\mbox{\boldmath $k$})}
%{4\alpha|\mbox{\boldmath $t$}
%(\varepsilon_{k\tau}^{*},\mbox{\boldmath $k$})|^3} 
%(\partial_{k_{x}}t_{x\tau}\partial_{k_{y}}t_{y\tau}
%-\partial_{k_{x}}t_{y\tau}\partial_{k_{y}}t_{x\tau}). \label{tah}
%\end{eqnarray}
As in the case of the anomalous Hall effect for the charge current,
the thermal anomalous Hall conductivity is also dominated by
the contributions from electrons occupying the momentum space
sandwiched between the spin-orbit-splitted Fermi surfaces.
This property brings about a remarkable effect
in superconducting states.
In the superconducting state, when
vortices are pinned in the mixed state, the Hall effect for the charge current
does not exist.
Instead, the thermal Hall effect for the heat current carried by
the Bogoliubov quasiparticles is possible.
Since we consider the magnetic field perpendicular to the $xy$-plane,
electrons in the normal core do not contribute to
the thermal transport in the direction parallel to the plane. 
Below the superconducting transition temperature
$\sigma_{xy}$ is infinite while $L^{(1)}$ is finite. 
Thus the thermal Hall effect is governed by
the first term of the right-hand side of (\ref{hh}), i.e.
the coefficient $L_{xy}^{(2)}$.
In contrast to the normal Hall effect for the heat current which 
decreases rapidly in the superconducting state, 
the coefficient $L_{xy}^{(2) {\rm AHE}}$
is not affected by the superconducting transition 
when the magnitude of the spin-orbit-splitting is much larger than
the superconducting gap as in the case of CePt$_3$Si and CeRhSi$_3$.  
Thus, even in the  limit of $T\rightarrow 0$, $\kappa^{\rm AHE}_{xy}/(H_zT)$
takes a finite value. 
Moreover in heavy fermion systems, 
the magnitude of $\kappa^{\rm AHE}_{xy}/(H_zT)$ 
in the limit of $T\rightarrow 0$ 
is expected to be much enhanced by the factor $\Lambda^{sz}$.

Finally, we discuss the validity of 
the disregard for the heat current carried by
the interaction between quasiparticles.
In the vicinity of the Fermi surface, the quasiparticle approximation
is applicable, and the interaction between quasiparticles is much
reduced by the wave function renormalization factor $z^2_{k\tau}$, and
may be negligible for heavy fermion systems. 
However, as seen from Eq.(\ref{tahll}), 
the thermal anomalous Hall conductivity is dominated by
the contributions away from the Fermi surface, which throws
a doubt on this approximation.
As a matter of fact, the above treatment of the heat current is justified
as far as the spin-orbit splitting 
$\alpha |\mbox{\boldmath $t$}(0,\mbox{\boldmath $k$})|$
is much smaller than the Fermi energy $E_F$.
For $\alpha |\mbox{\boldmath $t$}(0,\mbox{\boldmath $k$})|\ll E_F$, 
the quasiparticle approximation
is still valid for all electrons in the region sandwiched between
the spin-orbit splitted two Fermi surfaces.
The relation 
$\alpha |\mbox{\boldmath $t$}(0,\mbox{\boldmath $k$})|\ll E_F$ holds when
the carrier density is not so low. 
This condition is satisfied for any noncentrosymmetric heavy fermion
superconductors CePt$_3$Si, CeRhSi$_3$, CeIrSi$_3$, and UIr.

\subsubsection{Spin Hall effect}

\begin{figure}
\begin{center}
\includegraphics[width=6cm]{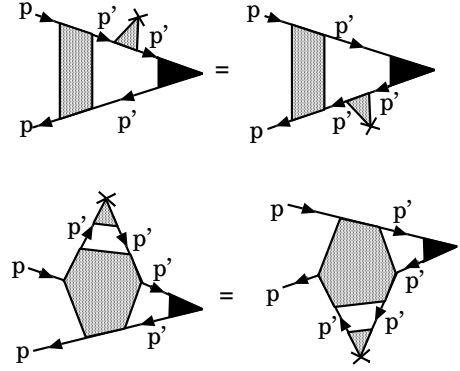}% Here is how to import EPS art
\end{center}
\caption{\label{fig:threevert} Diagrams representing
$\lim_{q\rightarrow 0}\nabla_{q}\hat{\mathcal{T}}^{\nu}(p+q/2,p-q/2)$.
The gray quadrangle is the four-point vertex irreducible with respect
to particle-hole pairs. The black triangle represents 
$\hat{\mathcal{T}}^{\nu}$.
The gray triangle with a cross
is a three-point vertex generated by $q$-derivative.
The gray hexagon is a six-point vertex irreducible with respect to
particle-hole pairs. } 
\end{figure}

Recently, the existence of the spin Hall effect
in the Rashba model has been extensively investigated by 
several authors.\cite{she,she2}
However, electron correlation effects on the spin Hall coefficient
has not yet been fully elucidated.
In this section, we explore the Fermi liquid theory for the spin Hall effect.
The spin Hall effect is characterized by 
the transverse spin current induced by
an electric field.
For the Rashba model with the inversion symmetry breaking along the $z$-axis, 
the in-plane spin current with the magnetization in the $z$-direction
is considered.
Then, the spin Hall conductivity is defined as, 
\begin{eqnarray}
\sigma^{\rm SHE}_{xy}=\lim_{\omega\rightarrow 0}
\frac{1}{i\omega}K^{\rm SHE}(i\omega_n)|_{i\omega_n\rightarrow \omega+i0},
\end{eqnarray}
\begin{eqnarray}
K^{\rm SHE}(i\omega_n)=
\int^{1/T}_0 d\tau \langle T_{\tau}\{J^{sz}_x(\tau)J_y(0)\}
\rangle e^{i\omega_n \tau}.
%\int^{\infty}_0 dt \langle [J^{sz}_x(t),J_y(0)]\rangle
%e^{i\omega t}.
\end{eqnarray}
Here the total spin current $J^{sz}_x$ is,
\begin{eqnarray}
J^{sz}_x=\frac{\mu_{\rm B}}{2}
\sum_kc^{\dagger}_k(\hat{v}_{kx}\sigma^z+
\sigma^z\hat{v}_{kx})c_k.
\end{eqnarray}
Note that we put the $g$ factor equal to 2.
The spin current vertex function fully-dressed by
electron-electron interaction is readily obtained from 
the Ward identity (\ref{swi}) which is, in the case of the Rashba interaction,
rewritten as,
\begin{eqnarray}
&&\sum_{\mu=0,x,y,z}q_{\mu}\hat{\Lambda}^{sz}_{\mu}(p+\frac{q}{2},p-\frac{q}{2})
= \nonumber \\
&&(-i\omega+\varepsilon_{k+\frac{q}{2}}+\Sigma_0(p+\frac{q}{2})
-\varepsilon_{k-\frac{q}{2}}-\Sigma_0(p-\frac{q}{2}))
\frac{\sigma^z}{2} \nonumber \\
&&-\frac{i\alpha}{2}\mbox{\boldmath $\sigma$}\cdot
[\mbox{\boldmath $t$}(p+\frac{q}{2})+
\mbox{\boldmath $t$}(p-\frac{q}{2})]
+\hat{\mathcal{T}}^z(p+\frac{q}{2},p-\frac{q}{2}). \label{swir}
\end{eqnarray}
We note that for $\omega/\mbox{\boldmath $q$}\rightarrow 0$ 
and small $\mbox{\boldmath $q$}$, $\hat{\mathcal{T}}^{\nu}(p+q/2,p-q/2)$ 
is expanded as,
\begin{eqnarray}
\hat{\mathcal{T}}^{\nu}(p+\frac{q}{2},p-\frac{q}{2})
=\hat{\mathcal{T}}^{\nu}(p,p)
+O(q^2). \label{tlexw}
\end{eqnarray}
The absence of the $q$-linear term in (\ref{tlexw})
is verified as follows.
For $\omega=0$,
in the Feynman diagrams which constitute $\hat{\mathcal{T}}^z(p+q/2,p-q/2)$, 
the total number of $G$-lines which depend on $q/2$ is equal to
that of $G$-lines which depend on $-q/2$.
Each diagram which is derived by differentiating a $G(p'+q/2)$-line
with respect to $\mbox{\boldmath $q$}$ gives the same contribution
as that obtained from the differentiation of $G(p''-q/2)$-line with respect to
$-\mbox{\boldmath $q$}$ in the limit of $q\rightarrow 0$.
This situation is diagrammatically depicted in FIG.\ref{fig:threevert}.
Thus, all terms of order $O(q)$ in
$q\cdot\nabla_q\hat{\mathcal{T}}^z(p+q/2,p-q/2)$
cancel with each other, which leads Eq.(\ref{tlexw}). 
Then,
putting $\omega=0$ first and then taking 
the limit of $\mbox{\boldmath $q$}\rightarrow 0$ in (\ref{swir}), we have
\begin{eqnarray}
\Lambda^{sz}_{\mu}(p,p)=\mu_{\rm B}
\frac{\partial}{\partial k_{\mu}}(\varepsilon_k
+\Sigma_0(p))\sigma^z. \label{spinc}
\end{eqnarray}
%In the derivation of (\ref{spinc}),
%we used Eq.(\ref{tlexw}).
We note that
in the expression of $\Lambda^{sz}_{\mu}$, the vertex corrections associated
with dissipative processes are not included, as in the
case of the anomalous Hall effect considered in Sec.II.D.1.
As long as the spin-orbit splitting is much larger than
the quasiparticle damping, such vertex corrections are not important.
%the fact that for small $q$,
%\begin{eqnarray}
%\hat{\mathcal{T}}^z(p+\frac{q}{2},p-\frac{q}{2})=\hat{\mathcal{T}}^z(p,p)
%+O(q^2), \label{tlex}
%\end{eqnarray}
%which follows from the same reason
%as explained below Eq.(tlexw).

By using (\ref{cv0}) and (\ref{spinc}), $K^{\rm SHE}(i\omega_n)$
is recast into,
\begin{eqnarray}
\sum_p{\rm tr}[\hat{\Lambda}^{sz}_x(p+q,p)\hat{G}(p)\hat{v}_{ky}
%\hat{\Lambda}^c_y(p,p+q)
\hat{G}(p+q)].
\end{eqnarray}
A straightforward calculation leads,
\begin{eqnarray}
\sigma^{\rm SHE}_{xy}=e\mu_{\rm B}\sum_k\sum_{\tau=\pm}
\frac{-\tau f(\varepsilon_{k\tau}^{*})
%\tanh(\frac{\varepsilon_{k\tau}^{*}}{2T})
\hat{t}_x(\varepsilon_{k\tau}^{*},\mbox{\boldmath $k$})\tilde{v}_{x\tau}
\partial_{k_{y}}t_{0y}}
{2\alpha |{\rm Re}~\mbox{\boldmath $t$}
(\varepsilon_{k\tau}^{*},\mbox{\boldmath $k$})|^2}
%\hat{t}_x(\varepsilon_{k\tau}^{*},\mbox{\boldmath $k$})\tilde{v}_{x\tau}
%\partial_{k_{y}}t_{0y}
, \label{she}
\end{eqnarray}
where 
\begin{eqnarray}
\tilde{v}_{\mu\tau}=\partial_{k_{\mu}}(\varepsilon_k+
\Sigma_0^R(\varepsilon,
\mbox{\boldmath $k$}))|_{\varepsilon=\varepsilon_{k\tau}^{*}}.
\end{eqnarray}
In the expression of the spin Hall conductivity (\ref{she}),
the factors $\tilde{v}_{x\tau}$ and $\mbox{\boldmath $t$}(p)$
are not substantially affected by electron 
correlation effects, because the main effect of the electron correlation
on these factors appears only through the deformation of the Fermi surface.
Thus, the spin Hall conductivity $\sigma_{xy}^{\rm SHE}$ is not renormalized
by electron correlation effects, but determined solely by the band structure.

\subsubsection{Magnetoelectric effect}

The existence of the spin-orbit interaction of the form 
$(\mbox{\boldmath t}_{0k}\times \nabla V)\cdot\mbox{\boldmath $\sigma$}$
gives rise to nontrivial coupling between charge and spin degrees of freedom,
which results in interesting magnetoelectric effects.
For example, there exists the bulk magnetization induced by
the charge current flow, as pointed out by Levitov {\it et al.} 
many years ago.\cite{lev}
To argue this phenomenon,
we consider the Rashba-type system with $\nabla V$ along the $z$-axis.
The magnetization along the $x$-direction is generated
by an electric field applied in the $y$-direction,
\begin{eqnarray}
M_x=\Upsilon_{xy}E_y.
\end{eqnarray}
The inverse effect is also possible; i.e.
an AC-magnetic field can yield the charge current flow,
\begin{eqnarray}
J_x=-2\Upsilon_{xy}\frac{\partial B_y}{\partial t}.\label{jb}
\end{eqnarray}
The prefactor $2$ in (\ref{jb}) is due to
the fact that in addition to the current directly induced by $\dot{B}$,
there is a magnetization current $c(\nabla\times M)$. 
It is noted that analogous effects in the superconducting state, i.e.
the supercurrent flow induced by the Zeeman magnetic field, and
its inverse effect, are also argued by Edelstein, Yip, and 
the present author.\cite{ede1,ede2,yip,fuji2}

The formula of the magnetoelectric coefficient $\Upsilon_{xy}$
in the case without electron-electron interaction was obtained by
Levitov {\it et al.}\cite{lev,ede0}
Here we consider effects of electron correlation on $\Upsilon_{xy}$,
which may be important in the application to heavy fermion superconductors.
In the normal state, the magnetoelectric coefficient $\Upsilon_{xy}$
is a non-equilibrium transport coefficient, to which
the Fermi liquid corrections are obtained by the method developed by
Eliashberg for the calculation of the electric conductivity.\cite{eli}

According to the Kubo formula, the magnetoelectric coefficient
is given by,
\begin{eqnarray}
\Upsilon_{xy}(\omega)=\frac{1}{i\omega}
K_{xy}^{\rm ME}(i\omega_n)|_{i\omega_n\rightarrow \omega+i0},
\end{eqnarray}
\begin{eqnarray}
K_{xy}^{\rm ME}(i\omega_n)=
\int^{1/T}_0 d\tau \langle T_{\tau}\{S^{x}(\tau)J_y(0)\}
\rangle e^{i\omega_n \tau}.
\label{mec}
\end{eqnarray}

Applying the Eliashberg's method,\cite{eli} we can 
express this correlation function in terms of
the Green functions and the three-point vertex functions for
the spin density and the charge current density.
The spin density vertex function required for
the calculation of (\ref{mec})
is obtained from
the Ward identity (\ref{swi}).
Taking the limit of $\mbox{\boldmath $q$}\rightarrow 0$ of the Ward identity
(\ref{swi}), we have the equation satisfied by 
the fully-dressed three-point vertex function of
the spin density,
\begin{eqnarray}
&&i\omega \hat{\Lambda}^{s\nu}_0(p+\frac{q}{2},p-\frac{q}{2})
=\frac{\sigma_{\nu}}{2}[i\omega-\Sigma_0(p+\frac{q}{2})
+\Sigma_0(p-\frac{q}{2})] \nonumber \\
&&-\frac{\alpha}{2}[\mathcal{L}_{\nu}(p+\frac{q}{2})-
\mathcal{L}_{\nu}(p-\frac{q}{2})]  \nonumber \\
&&+\frac{i\alpha}{2}(\{\mbox{\boldmath $\mathcal{L}$}(p+\frac{q}{2})
+\mbox{\boldmath $\mathcal{L}$}(p-\frac{q}{2})
\}\times\mbox{\boldmath $\sigma$})\cdot\mbox{\boldmath $n$}_{\nu}
\nonumber \\
&&-\hat{\mathcal{T}}^{\nu}(p+\frac{q}{2},p-\frac{q}{2}).
\label{sve}
\end{eqnarray}
In the above expression, $q=(i\omega,\mbox{\boldmath $0$})$.
%%%%%%%%%%%%%%%%%%amended
Using Eq.(\ref{tl}), we extract $\omega$-linear terms of the right-hand side of
(\ref{sve}) and take the limit $\omega\rightarrow 0$ arriving at,
%Using (\ref{tlexw}), we see that
%the third and the forth terms of the right-hand side of Eq.(\ref{sve})
%is
%\begin{eqnarray}
%i\alpha(\mbox{\boldmath $\mathcal{L}$}(p)\times
%\mbox{\boldmath $\sigma$})\cdot\mbox{\boldmath $n$}_{\nu}
%-\hat{\mathcal{T}}^{\nu}(p,p)+O(q^2)=O(q^2),
%\end{eqnarray}
%where we have used Eq.(\ref{tl}).
%As a result, the dressed three-point vertex function 
%for the spin density in the limit of 
%$\mbox{\boldmath $q$}/\omega\rightarrow 0$, $\omega\rightarrow 0$ is given by,
\begin{eqnarray}
\hat{\Lambda}^{s\nu}_0(p,p)=\frac{\sigma_{\nu}}{2}
\left(1-\frac{\partial \Sigma_0(p)}
{\partial (i\varepsilon_n)}\right)-\frac{\alpha}{2}
\frac{\partial \mathcal{L}_{\nu}(p)}{\partial (i\varepsilon_n)} \nonumber \\
-\frac{\partial \hat{\mathcal{T}}^{\nu}(p+q/2,p-q/2)}{\partial (i\omega)}
\biggr\vert_{i\omega\rightarrow 0}. 
\label{spinvertwt}
\end{eqnarray}
Then, using Eqs.(\ref{spinvertwt}) and (\ref{cv}) for
the spin density and charge velocity vertex functions, respectively,
and extracting the factor 
$G^R_{\tau}(p)G^A_{\tau}(p)=2\pi i 
z_{k\tau}^2\delta(\varepsilon-\varepsilon_{k\tau}^{*})
/(\omega+2i\gamma_{k\tau})$
after the analytical continuation to the real frequency,
we have for small $\omega$,
\begin{eqnarray}
\Upsilon_{xy}(\omega)=e\mu_{\rm B}
\sum_k\sum_{\tau=\pm}\frac{\tau z_{k\tau}^2\Pi_{k\tau}}
{4T\cosh^2(\frac{\varepsilon_{k\tau}^{*}}{2T})}\frac{i}
{\omega+2i\gamma_{k\tau}}, \label{me}
\end{eqnarray}
\begin{eqnarray}
\Pi_{k\tau}&=&\biggl(1-\frac{\partial \Sigma_0^R (\varepsilon_{k\tau}^{*},
\mbox{\boldmath $k$})}{\partial \varepsilon}
-\frac{\partial (\mathcal{T}^x_{\uparrow\downarrow}
+\mathcal{T}^x_{\downarrow\uparrow})}{2\partial \omega}
\biggr\vert_{\omega\rightarrow 0}\biggr)\mathcal{J}_{y\tau} \nonumber \\
&-&\tau\biggl(\frac{\partial 
\Sigma_{x}(\varepsilon_{k\tau}^{*},
\mbox{\boldmath $k$})}{\partial \varepsilon}
+\frac{\partial (\mathcal{T}^x_{\uparrow\uparrow}
+\mathcal{T}^x_{\downarrow\downarrow})}{2\partial \omega}
\biggr\vert_{\omega\rightarrow 0}
\biggr)\mathcal{I}_{y\tau},\label{pi}
\end{eqnarray}
with,
\begin{eqnarray}
\mathcal{J}_{y\tau}=\hat{t}_{y}\tilde{v}_{y\tau}+\tau\alpha\partial_{k_y}
t_{y\tau}+\Delta \mathcal{J},
\end{eqnarray}
\begin{eqnarray}
\mathcal{I}_{y\tau}=\tilde{v}_{y\tau}+\tau(\partial_{k_y}
t_{y\tau}\hat{t}_y+\partial_{k_y}
t_{x\tau}\hat{t}_x)+\Delta\mathcal{I}.
\end{eqnarray}
Here $\Delta\mathcal{J}$ and $\Delta\mathcal{I}$ are the contributions
from the vertex corrections to the charge current
which are required for preserving the momentum conservation law,
and related to the quasiparticle damping $\gamma_{k\tau}$
by the Ward identity for the charge degrees of freedom.\cite{yy}
In the following we are not concerned with the explicit form of these vertex 
corrections, but merely assume that umklapp processes or
impurity scattering break the momentum
conservation, leading to 
a finite value of the dissipative transport coefficient.

In the expression of Eq.(\ref{me}), the mass renormalization factor
$z_{k\tau}$ cancels after carrying out the momentum sum $\sum_k$
in the DC limit $\omega\rightarrow 0$,
and thus electron correlation effects appear only through
$\Pi_{k\tau}$ and
the factor $\gamma_{k\tau}/z_{k\tau}={\rm Im}\Sigma_{0}(\mbox{\boldmath $k$},
\varepsilon_{k\tau}^{*})$. 
This factor is related to
the resistivity $\rho \sim {\rm Im}\Sigma_{0}
\sim c_0+AT^2$, where
the constant $c_0$ is a residual resistivity due to impurity
scattering.
In heavy fermion systems, the coefficient of the $T$-square term of 
the resistivity $A$ is almost proportional to the square of
the specific heat coefficient $\gamma^2$, which are much enhanced
by electron correlation.
%%%%%%%%%%%amended 
The factor $\Pi_{k\tau}$ (\ref{pi}) is enhanced like the 
specific heat coefficient $\gamma$ because of the term
proportional to $-\partial \Sigma_0^R(p)/\partial \varepsilon$.
The other terms in $\Pi_{k\tau}$ are smaller than this term
by a factor $\alpha\vert\mbox{\boldmath $t$}(p)\vert/E_F\ll 1$.
%%%%%%%%%%%%%%%%%%%%%%
Thus,
\begin{eqnarray}
\Upsilon_{xy}\sim\frac{\gamma}{c_0+{\rm const.}\gamma^2T^2},
\end{eqnarray} 
at low temperatures.
In the temperature region where the resistivity exhibits $T^2$-dependence,
$\Upsilon_{xy}$ behaves like $\sim 1/(\gamma T^2)$, 
while in the zero temperature limit $\Upsilon_{xy}$ is enhanced by the factor
$\gamma$.

\section{Superconducting state}

\subsection{Basic equations}

The Fermi liquid theory in the normal state developed 
in the previous section can
be straightforwardly extended to the superconducting state
as in the case with inversion symmetry.\cite{leg,mig,fuji}
This task for the case of the Rashba interaction was done by the author 
in ref.\cite{fuji2}.
Here we present the formalism for general forms of the spin-orbit interaction
$\alpha \mbox{\boldmath $\mathcal{L}$}_0(k)
\cdot\mbox{\boldmath $\sigma$}$.
We add a pairing interaction to the model Hamiltonian (\ref{ham}),
which may be originated from higher order interaction processes caused by
the on-site Coulomb term $U$, or any other interactions.

In the conventional Nambu representation,\cite{sch}
the inverse of the single-particle
Green's function is defined as,
\begin{eqnarray}
&&\hat{\mathcal{G}}^{-1}(p)= \nonumber \\
&&\left(
\begin{array}{cc}
   i\varepsilon_n-\hat{H}(p)+\mu_{\rm B}\mbox{\boldmath $\sigma$}\cdot
\mbox{\boldmath $H$} & -\hat{\Delta}(p)   \\
   -\hat{\Delta}^{\dagger}(p) & i\varepsilon_n+\hat{H}^{t}(-p)
-\mu_{\rm B}\mbox{\boldmath $\sigma$}^{t}\cdot
\mbox{\boldmath $H$}
\end{array}
\right). \label{ginsc}
\end{eqnarray}
Here $\hat{H}(p)$ is given by the same expression as Eq.(\ref{h1}).
However, the self-energy $\hat{\Sigma}(p)$ in this case
includes both the normal Green function $\hat{G}(p)$ and
the anomalous Green function $\hat{F}(p)$ presented below
as internal lines.

$i\varepsilon_n-\hat{H}(p)+\mu_{\rm B}\mbox{\boldmath $\sigma$}\cdot
\mbox{\boldmath $H$}$ and $i\varepsilon_n+\hat{H}^t(-p)
-\mu_{\rm B}\mbox{\boldmath $\sigma$}^{t}\cdot
\mbox{\boldmath $H$}$ in 
$\hat{\mathcal{G}}^{-1}(p)$
are diagonalized by the transformation 
$\hat{\mathcal{A}}(p)\hat{\mathcal{G}}^{-1}(p)\hat{\mathcal{A}}_{+}(p)$
with,
\begin{eqnarray}
\hat{\mathcal{A}}(p)=
\left(
\begin{array}{cc}
\hat{U}(p) & 0 \\
 0 & \hat{U}^{t}_{+}(-p)
\end{array}
\right),
\end{eqnarray}
\begin{eqnarray}
\hat{\mathcal{A}}_{+}(p)=
\left(
\begin{array}{cc}
\hat{U}_{+}(p) & 0 \\
 0 & \hat{U}^{t}(-p)
\end{array}
\right),
\end{eqnarray}
Here $\hat{U}(p)$ and $\hat{U}_{+}(p)$ have 
the same forms as in the normal state given by 
(\ref{unita}) and (\ref{unita2}), respectively.

To diagonalize the superconducting gap $\hat{\Delta}(p)$,
we need the assumption that ${\rm Im}\Sigma_x$ and ${\rm Im}\Sigma_{y}$
(and thus ${\rm Im}\mathcal{L}_{x,y}$) are negligible, which implies
$\hat{U}_{+}(p)=\hat{U}^{\dagger}(p)$, and is justified
in the absence of magnetic fields,
%for any heavy fermion systems without inversion symmetry, in which
%$\alpha |\mbox{\boldmath 
%$\mathcal{L}$}(p)|\ll E_F$ holds.
%Also, it is required that there is no magnetic field, 
i.e. $\mbox{\boldmath $H$}=0$.
%In the absence of the magnetic field, $\mbox{\boldmath $H$}=0$,
%the superconducting 
%gap function $\hat{\Delta}(p)$ 
Under this condition, $\hat{\Delta}(p)$ and $\hat{\Delta}^{\dagger}(p)$
in $\hat{\mathcal{G}}^{-1}(p)$
are diagonalized by the transformation
$\hat{\mathcal{A}}\hat{\mathcal{G}}^{-1}(p)
\hat{\mathcal{A}}_{+}$ if and only if
the gap function has the form,
\begin{eqnarray}
\hat{\Delta}(p)=\Delta_s(k)i\sigma_y+\Delta_t(k)\mbox{\boldmath 
$\mathcal{L}$}(p)\cdot\mbox{\boldmath $\sigma$}i\sigma_y. \label{delt}
\end{eqnarray}
Eq.(\ref{delt}) implies that there is no Cooper pair between electrons on
different Fermi surfaces $\varepsilon_{k+}$ and $\varepsilon_{k-}$, of which 
the existence gives rise to pair-breaking effects.

The inverse of Eq.(\ref{ginsc}) is readily obtained as,
\begin{eqnarray}
\hat{\mathcal{G}}(p)=
\left(
\begin{array}{cc}
 \hat{G}(p)  & \hat{F}(p)   \\
  \hat{F}^{\dagger}(p) &  -\hat{G}^{t}(-p) 
\end{array}
\right), \label{g1}
\end{eqnarray}
where
\begin{eqnarray}
\hat{G}(p)=\sum_{\tau=\pm 1}
\frac{1+\tau\hat{\mbox{\boldmath $\mathcal{L}$}}(p)
\cdot\mbox{\boldmath $\sigma$}}{2}G_{\tau}(p),
\end{eqnarray}
\begin{eqnarray}
\hat{F}(p)=\sum_{\tau=\pm 1}
\frac{1+\tau\hat{\mbox{\boldmath $\mathcal{L}$}}(p)\cdot
\mbox{\boldmath $\sigma$}}{2}i\sigma_yF_{\tau}(p),
\end{eqnarray}
and, 
\begin{eqnarray}
G_{\tau}(p)=\frac{\Xi_{\tau}^{(-)}(p)}
{\Xi_{\tau}^{(+)}(p)\Xi_{\tau}^{(-)}(p)-\tilde{\Delta}_{k\tau}^2},
\label{bg1}
\end{eqnarray}
\begin{eqnarray}
F_{\tau}(p)=\frac{\tilde{\Delta}_{k\tau}}
{\Xi_{\tau}^{(+)}(p)\Xi_{\tau}^{(-)}(p)
-\tilde{\Delta}_{k\tau}^2},
\label{bg2}
\end{eqnarray}
\begin{eqnarray}
\Xi_{\tau}^{(\pm)}(p)=i\varepsilon\mp[\varepsilon_k-\mu+\Sigma_0(\pm p)+\tau
\alpha |\mbox{\boldmath $\mathcal{L}$}(\pm p)|],
\end{eqnarray}
%\begin{eqnarray}
%G_{\tau}(p)=\frac{i\varepsilon+\varepsilon_k-\mu+\Sigma_0(-p)+\tau
%\alpha |\mbox{\boldmath $\mathcal{L}$}(-p)|}
%{(i\varepsilon-\varepsilon_k+\mu-\Sigma_0(p)-\tau
%\alpha |\mbox{\boldmath $\mathcal{L}$}(p)|)
%(i\varepsilon+\varepsilon_k-\mu+\Sigma_0(-p)+\tau
%\alpha |\mbox{\boldmath $\mathcal{L}$}(-p)|)-\tilde{\Delta}_{k\tau}^2},
%\label{bg1}
%\end{eqnarray}
%\begin{eqnarray}
%F_{\tau}(p)=\frac{\tilde{\Delta}_{k\tau}}
%{(i\varepsilon-\varepsilon_k+\mu-\Sigma_0(p)-\tau
%\alpha |\mbox{\boldmath $\mathcal{L}$}(p)|)
%(i\varepsilon+\varepsilon_k-\mu+\Sigma_0(-p)+\tau
%\alpha |\mbox{\boldmath $\mathcal{L}$}(-p)|)-\tilde{\Delta}_{k\tau}^2},
%\label{bg2}
%\end{eqnarray}
with $\tilde{\Delta}_{k\tau}=\Delta_s(k)
+\tau|\mbox{\boldmath $\mathcal{L}$}(k)|\Delta_t(k)$.
When the quasiparticle approximation is applicable,
\begin{eqnarray}
G_{\tau}(p)=\frac{z_{k\tau}(i\varepsilon+\varepsilon_{k\tau}^{*})}
{(i\varepsilon+i\gamma_{k\tau} {\rm sgn}\varepsilon)^2-E^{2}_{k\tau}},
\label{scg}
\end{eqnarray}
\begin{eqnarray}
F_{\tau}(p)=\frac{z_{k\tau}\Delta_{k\tau}}
{(i\varepsilon+i\gamma_{k\tau} {\rm sgn}\varepsilon)^2-E^{2}_{k\tau}},
\label{scf}
\end{eqnarray}
with $E_{k\tau}=\sqrt{\varepsilon_{k\tau}^{*2}+\Delta^2_{k\tau}}$ and 
$\Delta_{k\tau}=z_{k\tau}\tilde{\Delta}_{k\tau}$.
The mass renormalization factor $z_{k\tau}$ is also given by
the same expression as 
Eq.(\ref{zkt}) with the self-energy $\hat{\Sigma}(p)$ replaced
with that in the superconducting state.

Occasionally, some physical quantities for inversion-symmetry-broken 
systems are governed by electrons occupying 
the momentum space region between the spin-orbit splitted Fermi surfaces,
as argued in the previous sections.
In such cases, when the spin-orbit splitting is sufficiently large, 
the quasiparticle approximation cannot be generally justified,
and it may be required to use the complicated expressions Eqs.(\ref{bg1})
and (\ref{bg2}).
However, fortunately, for the noncentrosymmetric heavy fermion 
superconductors CePt$_3$Si and CeRhSi$_3$, 
the spin-orbit splitting is much larger than
the superconducting gap, and thus any physical quantities
governed by electrons occupying 
the momentum space region between the spin-orbit splitted Fermi surfaces
are safely approximated by the quantities in the normal state,
of which the calculation is much easier.

\subsection{Paramagnetic properties}

We now argue the paramagnetism of this system on the basis of
the formalism presented above.
The uniform spin susceptibility is calculated from
the $\nu$-component of the magnetization 
\begin{eqnarray}
M_{\nu}=\frac{T}{2}\sum_{p}{\rm tr}\left[\left(
\begin{array}{cc}
\sigma_{\nu} & 0 \\
0 & -\sigma_{\nu}^{t} \\
\end{array}\right)
%M_{\nu}=\frac{T}{2}\sum_{p}{\rm tr}[\tau_z\otimes\sigma_{\nu}
\hat{\mathcal{G}}(p)\right].
\end{eqnarray}
%with $\tau_z$ the $z$-component of the Pauli matrix.
In the derivation of the spin susceptibility, we ignore
the magnetic-field-dependence of the superconducting gap,
because 
the contributions from $\partial \Delta_{k\tau}/\partial H_{\nu}$
to the spin susceptibility is vanishingly small
when the energy dependence of the
density of states is nearly particle-hole symmetric.
 
In the case with cubic symmetry, 
the uniform spin susceptibility in the superconducting state
is calculated as,
\begin{eqnarray}
&&\chi_{zz}(T)=\mu_{\rm B}^2\sum_{\tau=\pm}\sum_k\frac{z_{k\tau}}{4T
\cosh^2(\frac{E_{k\tau}}{2T})}
\Lambda_{\rm P}^{\rm cub}(E_{k\tau},\mbox{\boldmath $k$}) 
\nonumber \\
&&-2\mu_{\rm B}^2T\sum_{\varepsilon_n}\sum_k
[G_{+}(p)G_{-}(p)+F_{+}(p)F_{-}(p)] \nonumber \\
&&\times
\Lambda^{\rm cub}_{\rm V}
(i\varepsilon_n,\mbox{\boldmath $k$}).
\label{sccu}
\end{eqnarray}
The first and second terms of the right-hand side of (\ref{sccu})
are, respectively, the Pauli and van-Vleck-like contributions.
The three-point vertex functions $\Lambda^{\rm cub}_{\rm P}$
and $\Lambda^{\rm cub}_{\rm V}$ in Eq.(\ref{sccu}) are 
slightly different 
from those appeared in the normal state 
(\ref{tv1}) and (\ref{tv2}) because the anomalous Green function $\hat{F}(p)$
also contributes to the self-energy $\hat{\Sigma}(p)$.
However, as long as there is no strong spin fluctuation,
the corrections to $\Lambda^{\rm cub}_{\rm P,V}$ due to
the existence of the superconducting gap are small.\cite{leg}
In the case of $|\Delta_{k\tau}|\ll\alpha |\mbox{\boldmath $\mathcal{L}$}(p)|$,
the quasiparticle approximation 
(\ref{scg}) and (\ref{scf}) is not applicable to the van-Vleck-like term.
Thus, we will not write down its explicit form in Eq.(\ref{sccu}).
As a matter of fact, in this case,
the van Vleck term is well approximated by the formula for the normal state
(\ref{chicu}). 
In the case of a spherical Fermi surface without
electron-electron interaction, and 
$\mbox{\boldmath $\mathcal{L}$}_{0}(k)=(k_x(k_y^2-k_z^2),
k_y(k_z^2-k_x^2),k_z(k_x^2-k_y^2)$,
if the condition 
$|\Delta_{k\tau}|\ll \alpha |\mbox{\boldmath $\mathcal{L}$}(p)|
\ll E_F$ is satisfied, the spin susceptibility at zero temperature
is $\chi_{zz}(0)\approx \frac{2}{3}\chi_{zz}(T_c)$, since
the van-Vleck-like term remains finite.

In the Rashba case, for a magnetic field perpendicular to
the plane, 
the uniform spin susceptibility is given by,
\begin{eqnarray}
\chi_{zz}(T)&=&-2\mu_{\rm B}^2T\sum_{\varepsilon_n}\sum_k
[G_{+}(p)G_{-}(p)+F_{+}(p)F_{-}(p)] \nonumber \\
&&\times
\Lambda^{sz}(i\varepsilon_n,\mbox{\boldmath $k$}). \label{scchizz}
\end{eqnarray}
For an in-plane magnetic field,\cite{fuji3}
\begin{eqnarray}
\chi_{xx}(T)&=&\mu_{\rm B}^2\sum_{\tau=\pm}\sum_k\frac{z_{k\tau}}{4T
\cosh^2(\frac{E_{k\tau}}{2T})}
\hat{t}_y\Lambda_{\tau}^{sx}(E_{k\tau},\mbox{\boldmath $k$}) \nonumber \\
&-&2\mu_{\rm B}^2T\sum_{\varepsilon_n}\sum_k
[G_{+}(p)G_{-}(p)+F_{+}(p)F_{-}(p)] \nonumber \\ 
&&\times\hat{t}_x\Lambda^{sx}_{+-}(i\varepsilon_n,\mbox{\boldmath $k$}).
\label{scchi}
\end{eqnarray}
Here we have ignored the corrections to the three-point vertex functions
due to the superconducting gap again assuming
that there is no strong ferromagnetic fluctuation.
We are concerned with the case that the typical size of the
spin-orbit splitting $2\alpha |\mbox{\boldmath $t$}(p)|$
 is much larger than $|\Delta_k|$; i.e. the situation relevant to
 CePt$_3$Si and CeRhSi$_3$.
In this case, the second term of the right-hand side of 
(\ref{scchi}), which is the van-Vleck-like contribution,
is approximated by the expression for the normal state 
$\chi_{xx}^{\rm VV}$ (\ref{chixv}), and not affected by
the superconducting transition, while
the Pauli term, i.e. the first term of (\ref{scchi}),
vanishes at zero temperature.
When the energy dependence of the density of states is sufficiently small,
e.g. the case of the spherical Fermi surface $\varepsilon=k^2/(2m)$,
we have $\chi_{xx}^{\rm SC}(T=0)/\chi_{xx}^{\rm SC}(T=T_c)=1/2$,
as pointed out by Edelstein, and Gorkov and Rashba.\cite{ede1,gor}
We would like to stress that the nonzero $\chi$ at zero temperature
in the superconducting state is not due to the admixture
of the spin singlet and triplet pairs, but rather
attributed to the existence of the van-Vleck-like term
caused by the spin-orbit interaction.

\begin{figure}
\begin{center}
\includegraphics[width=3.5cm]{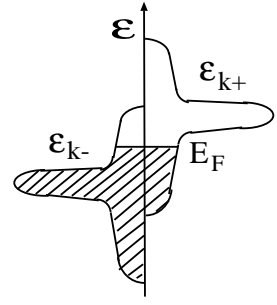}% Here is how to import EPS art
\end{center}
\caption{\label{fig:dossche} A schematic figure of 
an example of the density of states
which is significantly enhanced in the energy region sandwiched
by the spin-orbit splitted two bands $\varepsilon_{k+}$ and 
$\varepsilon_{k-}$.} 
\end{figure}

\begin{figure}
\begin{center}
\includegraphics[width=7.8cm]{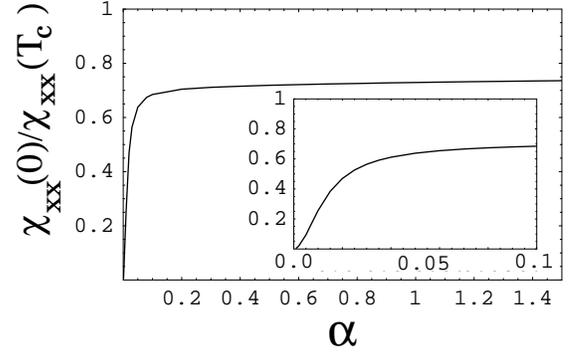}% Here is how to import EPS art
\end{center}
\caption{\label{fig:chira} $\chi_{xx}(T=0)/\chi_{xx}(T=T_c)$ 
in the superconducting state versus 
$\alpha$ for the model defined by Eqs.(\ref{ham}), (\ref{sq}), and (\ref{sqra})
with $U=0$. 
Inset: the enlarged plot for small $\alpha$} 
\end{figure}

\begin{figure}
\begin{center}
\includegraphics[width=6cm]{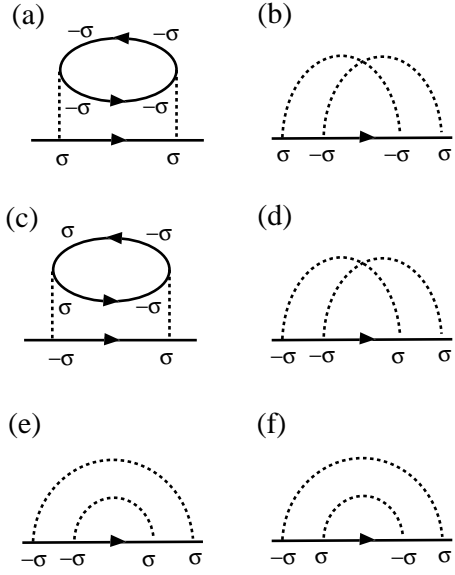}% Here is how to import EPS art
\end{center}
\caption{\label{fig:selfene} 
Diagrams for the self-energy  on the order of $U^2$.
Up to this order, the diagram (a) gives the most prominent
electron correlation effects.
The contributions from the other diagrams are suppressed by
factors of order  
$\sim O(\alpha |\mbox{\boldmath $t$}_{0k}|/E_F)$}. 
\end{figure}

\begin{figure}
\begin{center}
\includegraphics[width=7cm]{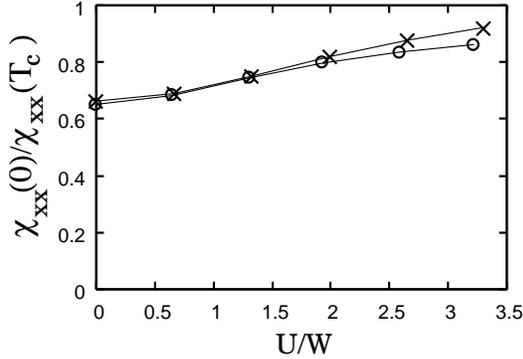}% Here is how to import EPS art
\end{center}
\caption{\label{fig:eechira} $\chi_{xx}(T=0)/\chi_{xx}(T=T_c)$ 
in the superconducting state versus 
$U/W$ with $\alpha=0.5$ (open circles) and $\alpha=1.0$ (crosses) calculated 
by the second order perturbative expansion with respect to $U$
for the model defined by Eqs.(\ref{ham}), (\ref{sq}), and (\ref{sqra}). 
} 
\end{figure}

\subsection{Implication for the Knight shift measurement of CePt$_3$Si}

Here we would like to argue an implication of (\ref{scchi})
for CePt$_3$Si.
There are strong pieces of experimental evidence suggesting
the realization of 
an unconventional pairing state in CePt$_3$Si.\cite{yogi,yogi2,matsu,pene}
Nevertheless, the symmetry of the Cooper pair has not yet
been clarified so far.
According to the Knight shift measurements done 
by Yogi {\it et al.}\cite{yogi,yogi2} 
%and
%the $\mu$SR experiments carried out  by Higemoto et al. 
for this system, 
both $\chi_{xx}$ and $\chi_{zz}$ 
does not show any significant change even below 
the superconducting transition temperature $T_c$.
The absence of the substantial decrease of $\chi_{xx}$ for $T<T_c$
seems not to be consistent with Eq.(\ref{scchi}), which indicates
the decrease of the Pauli term
both in the spin singlet and triplet superconducting states.
To resolve this point, we note that
the magnitude of the ratio
$\chi_{xx}(T=0)/\chi_{xx}(T=T_c)$ in the superconducting state 
crucially depends on
the details of the electronic structure.
For instance, when the density of states has the strong energy-dependence
as shown in FIG.7, the value of 
$\chi_{xx}(T=0)/\chi_{xx}(T=T_c)$ may
deviate from 1/2.
To demonstrate this, we use the model considered in Sec.II.C., 
given by the Hamiltonian (\ref{ham}) with Eqs.(\ref{sq}) and (\ref{sqra})
at the half-filling.
This model has the density of states similar to that depicted in
FIG.7 because of the existence of the van Hove singularity.
We chose the magnitude of the superconducting gap as
$\Delta_{\pm}\approx 0.01$, and calculate 
the spin susceptibility at zero temperature
from Eq.(\ref{scchi}).
In FIG.8, we show the plot of
 $\chi_{xx}(T=0)/\chi_{xx}(T=T_c)$ 
versus the strength of the spin-orbit interaction $\alpha$
for the non-interacting case $U=0$.
It is seen that, for small $\alpha$, 
$\chi_{xx}(T=0)/\chi_{xx}(T=T_c)$ 
is nearly equal to zero. As $\alpha $ increases,
the value of the ratio becomes larger, and eventually
goes over $1/2$, 
because of the enhancement of the density of states
due to the van Hove singularity.
The maximum value of the ratio is still
smaller than the experimentally observed value for CePt$_3$Si, i.e. 
$\chi_{xx}(T=0)/\chi_{xx}(T=T_c)\approx 1$.
However, it is expected that if electron
correlation effects are included, the van-Vleck-like term enjoys
the enormous enhancement caused by the contributions from
the van Hove singularity, which
is absent for the Pauli term if the spin-orbit splitted Fermi surfaces are
sufficiently away from the van Hove points.
To confirm this prediction, we carry out
the perturbative calculation of the electron
correlation effects by expanding
the self-energy in terms of the Coulomb interaction $U$
up to the second order.
We consider the case where the spin-orbit
splitting is much larger than the superconducting gap.
Then, $\chi_{xx}(T=0)$ can be well approximated 
by the van-Vleck-like susceptibility in the normal state.
Using the expressions in the normal state Eqs.(\ref{chixp}) and
(\ref{chixv}), we compute numerically
the ratio $\chi_{xx}(T=0)/\chi_{xx}(T=T_c)$ for $\alpha=0.5$ and $\alpha=1.0$.
Since the spin-orbit splitting is still much smaller than the band width
$W\approx 8$,
we neglect corrections of order $O(\alpha |\mbox{\boldmath $t$}_{0k}|/W)$, 
preserving only the contribution
from the diagram FIG.9(a).
This amounts to
considering only the term proportional to 
$\partial \Sigma_x/\partial (\mu_{\rm B}H_x)$ 
in Eqs.(\ref{svert2}) and (\ref{svert3}).
Since the numerical
computation of the derivative of the self-energy with respect to
a magnetic field is somewhat difficult, we use the approximation
$\partial  \Sigma_x/\partial (\mu_{\rm B}H_x)\approx 
\partial  \Sigma/\partial (i\varepsilon)$, which is
justified when a particular mode of spin fluctuation is not
developed, as in the case of CePt$_3$Si.\cite{yy}
The calculated results of the ratio plotted against $U/W$
are shown in FIG.10. 
Here $W$ is defined by the average of the band widths of
$\varepsilon_{k+}$ and $\varepsilon_{k-}$.
As $U/W$ increases, the enhancement of the van-Vleck-like susceptibility
due to electron correlation
overcomes that of the Pauli term, and the ratio
$\chi_{xx}(T=0)/\chi_{xx}(T=T_c)$ approaches 1.0.
Although the strength of the electron-electron interaction 
for which the ratio is nearly equal to 1.0
is considerably large, 
we believe that the result obtained by the perturbative calculation
may be qualitatively correct, and will be improved by including higher order
corrections as long as the system is in the Fermi liquid state.
Actually, CePt$_3$Si is a Fermi liquid with the mass enhancement factor
of order $\sim 60$, in which the spin susceptibility is also
substantially enhanced.
Since we cannot derive such a large mass enhancement
from ab initio calculations, we regard the parameter $U$ as an
effective interaction renormalized by correlation effects
in the above calculations.

Thus, it is possible that for CePt$_3$Si, the absence of the
decrease of $\chi_{xx}(T)$ below $T_c$ observed by 
the NMR experiment
may be attributed to
a specific energy dependence of the density of states, which leads
the existence of the van-Vleck-like term much
larger than the Pauli term.
Unfortunately, 
the LDA calculations of the electronic structure of this system
done by some groups are restricted within the case without 
the antiferromagnetic phase transition, which occurs at $T_N\sim 2 {\rm K}$
in CePt$_3$Si.\cite{hari,sam} 
Because of this reason, it is difficult to
clarify from first principle calculations
whether such a strong energy dependence of  the density of
states exists or or not for CePt$_3$Si.
Experimental studies which directly probe the energy dependence of 
the density of states
for this system are desirable.

\subsection{Magnetoelectric effect: Paramagnetic supercurrent}

In the absence of inversion symmetry, 
an intriguing magnetoelectric effect 
exists in the superconducting state as well as in the normal state, 
as extensively argued by
Edelstein, Yip, and the present author.\cite{ede1,ede2,yip,fuji2}
In the case of the Rashba interaction with a potentail gradient
along $\mbox{\boldmath $n$}=(0,0,1)$,
an in-plane magnetic field $\mbox{\boldmath $B$}$
induces a paramagnetic supercurrent,
$\mbox{\boldmath $J$}=\mathcal{K}(\mbox{\boldmath $n$}\times
\mbox{\boldmath $B$})$.
Here $\mathcal{K}$ is the magnetoelecric effect coefficient,
of which the explicit expression including electron correlation effects
is presented in ref.\cite{fuji2}.
The inverse effect is also possible; i.e.
the diamagnetic supercurrent flow gives rise to
a magnetization.\cite{ede2}
In the case of the Dresselhaus interaction with
$\mbox{\boldmath $\mathcal{L}$}_{0}(k)=(k_x(k_y^2-k_z^2),
k_y(k_z^2-k_x^2),k_z(k_x^2-k_y^2))$,
a similar effect exists, but
the paramagnetic supercurrent flows parallel to the applied magnetic field,
$\mbox{\boldmath $J$}=\mathcal{K}
\mbox{\boldmath $B$}$,
in contrast to the Rashba case where
the current flows perpendicular to the magnetic field. 
The magnetoelectric coefficient $\mathcal{K}$ is also directly related
to the stability of the helical vortex phase considered by Kaur
{\it et al.}, and determines the magnitude of the center of mass momentum
for inhomogeneous Cooper pairs.\cite{kau}

Recently,
it was pointed out by Yip that 
the paramagnetic supercurrent is exactly canceled with
the magnetization current $\mbox{\boldmath $J$}_M=c\nabla\times 
\mbox{\boldmath $M$}$ in the complete Meissner state
in a semi-infinite system.\cite{yip2}
However, in finite systems, this cancellation is imperfect, and
thus the paramagnetic supercurrent gives nonzero contributions
to the bulk current.
Moreover, in the mixed state, 
the cancellation is not complete generally.
Here we would like to re-examine electron correlation effects on
the paramagnetic supercurrent
taking into account the partial cancellation with the magnetization current
in the case of the Rashba interaction.

Following Yip, we start from the following relations
for the supercurrent density $\mbox{\boldmath $J$}_s$ and the magnetization
density $\mbox{\boldmath $M$}$,\cite{yip2}
\begin{eqnarray}
\mbox{\boldmath $J$}_s=\frac{1}{2e\Lambda}(\hbar\nabla\phi-\frac{2e}{c}
\mbox{\boldmath $A$})+\mathcal{K}(\mbox{\boldmath $n$}\times
\mbox{\boldmath $B$}), \label{scab}
\end{eqnarray}
\begin{eqnarray}
\mbox{\boldmath $M$}=\frac{\mathcal{K}}{2e}[\mbox{\boldmath $n$}\times
(\hbar\nabla\phi-\frac{2e}{c}
\mbox{\boldmath $A$})]+\mbox{\boldmath $M$}_{\rm Zee}.
\label{mab}
\end{eqnarray}
Here the first term of the right-hand side of
Eq.(\ref{scab}) is the diamagnetic supercurrent density which is denoted
by $\mbox{\boldmath $J$}^{\rm dia}=(J_x^{\rm dia},J_y^{\rm dia},
J_z^{\rm dia})$ in the following.
The second term of the right-hand side of (\ref{scab})
is the paramagnetic supercurrent induced by
the inversion-symmetry-breaking spin-orbit interaction.
The coefficient $1/\Lambda$ of the diamagnetic supercurrent
is related to the superfluid density,
and equivalent to the Drude weight at zero temperature.
The first term of the right-hand side of Eq.(\ref{mab})
is the magnetization induced by the magnetoelectric effect, and
the second term $\mbox{\boldmath $M$}_{\rm Zee}$
is a magnetization caused by the usual Zeeman effect.
For small $\mbox{\boldmath $B$}$, 
$\mbox{\boldmath $M$}_{\rm Zee}=\chi \mbox{\boldmath $B$}$. 
Eqs.(\ref{scab}) and (\ref{mab}) are obtained from
the free energy in the absence of inversion symmetry
as argued by several authors.\cite{yip2,ede3,sam2,kau}

In addition to the supercurrent $\mbox{\boldmath $J$}_s$, there exists
the magnetization current density $\mbox{\boldmath $J$}_M=c\nabla\times 
\mbox{\boldmath $M$}$,
which is derived by
adding the term 
\begin{eqnarray}
-\int d\mbox{\boldmath $r$}\mbox{\boldmath $M$}\cdot\mbox{\boldmath $B$}
=-\int d\mbox{\boldmath $r$}
(\nabla\times\mbox{\boldmath $M$})\cdot\mbox{\boldmath $A$}
\end{eqnarray}
to the free energy.
Using Eqs.(\ref{scab}), (\ref{mab}), and
the relation $\nabla\times \mbox{\boldmath $J$}^{\rm dia}=
-\mbox{\boldmath $B$}/(c\Lambda)$, we have\cite{vc}
\begin{eqnarray}
&&\mathcal{K}(\mbox{\boldmath $n$}\times
\mbox{\boldmath $B$})= -c\nabla\times(\mbox{\boldmath $M$}
-\mbox{\boldmath $M$}_{\rm Zee})  \nonumber \\
&&+c\mathcal{K}\Lambda(-\partial_xJ_z^{\rm dia},\partial_yJ_z^{\rm dia},
\partial_xJ_x^{\rm dia}+\partial_yJ_y^{\rm dia}).
\end{eqnarray}
Then, the total current density $\mbox{\boldmath $J$}_{\rm tot}$
is recast into
\begin{eqnarray}
\mbox{\boldmath $J$}_{\rm tot}&=&
\mbox{\boldmath $J$}_s+\mbox{\boldmath $J$}_M \nonumber \\
&=&\mbox{\boldmath $J$}^{\rm dia}+c\nabla\times\mbox{\boldmath $M$}_{\rm Zee}
\nonumber \\
&+&c\mathcal{K}\Lambda(-\partial_xJ_z^{\rm dia},\partial_yJ_z^{\rm dia},
\partial_xJ_x^{\rm dia}+\partial_yJ_y^{\rm dia}). \label{jt}
\end{eqnarray}
The second term of the right-hand side of Eq.(\ref{jt})
is the magnetization current caused by
the usual Zeeman effect.
The last term of Eq.(\ref{jt})
is the paramagnetic supercurrent due to the magnetoelectric effect with which
we are concerned.
Apparently, this paramagnetic term vanishes in the thermodynamic limit
in the complete Meissner state in
an infinite $x$-$y$ planar system, as was pointed out by Yip.\cite{yip2}
Nevertheless, it still plays a crucial role
in finite systems in the Meissner state, 
or generically in the mixed state.  Using Eq.(\ref{jt}), 
we can discuss electron correlation effects
on the paramagnetic supercurrent partly 
canceled with the magnetization current.
According to the analysis presented in ref.\cite{fuji2,add},
$\mathcal{K}$ is given by
\begin{eqnarray}
&&\frac{\mathcal{K}}{e\mu_{\rm B}}= \nonumber \\
&&\sum_k\sum_{\tau=\pm 1}\tau v_{0y\tau}
\frac{z_{k\tau}\Delta_{k\tau}^2}{E_{k\tau}^{2}}
\left[\frac{{\rm ch}^{-2}\frac{E_{k\tau}}{2T}}{2T}-
\frac{{\rm th}\frac{E_{k\tau}}{2T}}{E_{k\tau}}\right]
\Lambda^{sx}_{\tau}(E_{k\tau},\mbox{\boldmath $k$}) 
\nonumber \\
&&+2\alpha\sum_k\frac{\Delta_{k+}\Delta_{k-}}
{E_{k+}^2-E_{k-}^2}\left[z_{k-}\frac{{\rm th}\frac{E_{k+}}{2T}}{E_{k+}}
-z_{k+}\frac{{\rm th}\frac{E_{k-}}{2T}}{E_{k-}}\right] \nonumber \\
&&\times
\hat{t}_x\Lambda^{sx}_{+-}(E_{k\tau},\mbox{\boldmath $k$}), \label{kyx}
\end{eqnarray}
which is enhanced by the factor $1/z_{k\tau}$, when
the Wilson ratio is nearly equal to 2, as in the case of 
typical heavy fermion systems.
%According to ref.\cite{fuji2}, the magnetoelectric coefficient 
%$\mathcal{K}$ is not affected by electron correlation effects,
%while the coefficient of the diamagnetic supercurrent $1/\Lambda$
%is suppressed by the mass renormalization factor $\sim z_{k\tau}$.
On the other hand, 
the coefficient of the diamagnetic supercurrent $1/\Lambda$
is suppressed by the factor $z_{k\tau}$.
This fact indicates that
the last term of the right-hand side of 
(\ref{jt}) is enhanced by the factor $1/z_{k\tau}^2$ compared to
the first term $\mbox{\boldmath $J$}^{\rm dia}$.
Thus, even when
the partial cancellation with the magnetization current takes place.
the paramagnetic supercurrent induced by the magnetoelectric effect
is still amplified in strongly correlated electron systems with
large mass enhancement.
This observation is quite important for
the experimental detection of the magnetoelectric effect
in heavy fermion superconductors such as CePt$_3$Si, CeRhSi$_3$,
UIr, and CeIrSi$_3$.

\section{Summary}

In this paper, we have explored magnetic properties and transport
phenomena in interacting electron systems without inversion symmetry
in the normal state as well as in the superconducting state 
on the basis of the general
framework of the Fermi liquid theory, taking into account
electron correlation effects in a formally exact way.
We have obtained the formulae for the transport coefficients
related to the anomalous Hall effect, the thermal 
anomalous Hall effect, the spin Hall effect, and the magnetoelectric effect,
which are the results of the parity violation.
It is found that the spin Hall conductivity is not renormalized
by electron correlation effects, and is determined solely by
the band structure.
Also, it is pointed out that 
in the superconducting state, 
the thermal anomalous anomalous Hall conductivity divided by
the temperature and the magnetic field $\kappa_{xy}^{\rm AHE}/(TH_z)$
remains finite even in the zero temperature limit, despite the 
absence of Bogoliubov quasiparticles carrying heat currents.
This intriguing property is due to the fact that
the anomalous Hall conductivity in the absence of inversion symmetry 
is dominated by electrons
occupying the momentum space sandwiched between
the spin-orbit splitted two Fermi surfaces, which are not affected
by the superconducting transition.
We have also examined electron correlation effects on
the paramagnetic supercurrent caused by the magnetoelectric effect
taking account of the partial cancellation
with the magnetization current.
The experimental detection of these effects in
the recently discovered noncentrosymmetric heavy fermion superconductors
CePt$_3$Si, UIr, CeRhSi$_3$, and CeIrSi$_3$.
is an important future issue.

Furthermore, we have demonstrated that in the normal state
the temperature dependence of
the van-Vleck-like spin susceptibility behaves like the Pauli susceptibility,
in contrast to the usual van Vleck orbital
susceptibility, and that
the ratio of the van-Vleck-like term to the Pauli term
depends crucially on the details of the electronic structure
and electron correlation effects.
This result leads us a possible explanation for
the recent NMR experimental data of CePt$_3$Si, which indicates
no change of the Knight shift below $T_c$
for any directions of a magnetic field, contrary to
previous theoretical predictions.\cite{yogi2}
It is suggested that
the strong energy dependence of the density of states and
electron correlation effects may strongly enhance the van-Vleck-like
susceptibility compared to the Pauli term, and thus
the total spin susceptibility is not much affected by
the superconducting transition.

\acknowledgments{}

The author would like to thank K. Yamada, Y. Onuki,
Y. Matsuda, T. Shibauchi, H. Mukuda, M. Yogi, N. Kimura, T. Takeuchi,
and H. Ikeda for invaluable discussions.
He is also grateful to S. K. Yip for discussions about
the cancellation of the paramagnetic supercurrent.
The numerical calculations are performed on SX8 at YITP
in Kyoto University.
This work was partly supported by a Grant-in-Aid from the Ministry
of Education, Science, Sports and Culture, Japan.

%\begin{references}

%\end{references}


\begin{thebibliography}{99}
\bibitem{fuji3} The expression of $\chi_{xx}$ given in ref.\cite{fuji2}
should be amended as Eq.(\ref{scchi}) in the present paper.
\bibitem{vc} In the case of the mixed state, naturally, this argument is
applied only to the region outside of vortex cores.
\bibitem{add} In ref.\cite{fuji2}, the final expression of 
$\mathcal{K}$ (Eq.(36) in ref.\cite{fuji2}) should be amended
as Eq.(\ref{kyx}) in the present paper, because, in ref.\cite{fuji2}, 
the quasiparticle approximation
was erroneously applied to the energy region away from the Fermi level.
This correction 
does not flaw the main conclusion of ref.\cite{fuji2}, but rather
justifies its validity.
\end{thebibliography}
\end{document}